\documentclass[13pt,english]{article}

\usepackage[utf8]{inputenc}
\usepackage{cmap}
\usepackage[T1]{fontenc}
\usepackage{amsmath,amssymb,amstext}
\usepackage{babel}
\usepackage{graphicx}
\usepackage{xcolor}
\usepackage{times}
\usepackage{geometry}
\usepackage{hyperref}

\begin{document}

\title {Dynamics of stress propagation in anharmonic crystals: MD simulations.}  

\author{Zbigniew Kozioł\footnote{zbigniew.koziol@ncbj.gov.pl}, \\
	\small{National Center for Nuclear Research, Materials Research Laboratory},\\
	\small{ul. Andrzeja Sołtana 7, 05-400 Otwock-Świerk, Poland}}

\maketitle

\begin{abstract}
Anharmonic inter-atomic potential $\sim |x|^n$, $n>1$, has been used in molecular dynamics (MD) simulations of stress dynamics of FCC oriented crystal. 
The model of the chain of masses and springs is found as a convenient and accurate description of simulation results, with masses representing the crystallographic planes. The dynamics of oscillations of two planes is found analytically to be given by Euler's beta functions, and its scaling with non-linearity parameter and amplitude of oscillations, or applied shear pressure is discussed on examples of time dependencies of displacements, velocities, and forces acting on masses (planes). The dynamics of stress penetration from the surface of the sample with multiply-planes (an anharmonic crystal) towards its interior is confirmed to be given exactly as a series of Bessel functions, when n=2 (Schr\"{o}dinger and Pater solutions). When n $\ne$ 2 the stress dynamics (wave propagation) in bulk material remains qualitatively of the same nature as in the harmonic case. In particular, results suggest that the quasi-linear relationship between frequency and the wave number is preserved. The speed of the transverse sound component, dependent on sound wave amplitude, is found to be a strongly decreasing function of $n$. The results are useful in the analysis of any MD simulations under pressure, as they help to understand the dynamics of pressure retarded effects, as well as help design the proper methodology of performing MD simulations in cases such as, for instance, studies of the dynamics of dislocations. 
\end{abstract}

\vspace{2pc}

\section{Introduction.}
\label{Intro}

The proper understanding of the dynamics of crystalline solids is a key aspect of materials science. Tools and techniques for visualization of such dynamics are in general lacking \cite{PNAS}. As a consequence, materials models have suffered from a lack of input from experiments. A partial remedy to this is MD simulations, providing insight into sub-picosecond time scales and subatomic spatial resolution. 

There are several questions we found when performing MD simulations. For instance, how to define (strictly, in physical terms) the stress-strain region when crystal response remains linear? It appears that MD results provide a description of a dynamic process when a pressure is imposed on a crystal and there is no simple method developed that would allow extracting the linear response from simulation data. On another hand, how to find out the value of pressure acting on a dislocation in material, while the penetration process is a dynamic one of a comparable timescale as the time of simulation itself? Additionally, during the pressure front passage through the dislocation, interactions between atoms are strongly non-linear, and energy losses occur. Questions like these inspired us to find a description of the dynamics of atoms in crystals in a case when a strongly nonlinear, well-defined inter-atomic potential is used. We choose it in the form $|x|^n$, with $n>1$, since it is given then by a simple analytical function, and in the special case of $n=2$ it allows a comparison with the most common, classical situation of harmonic dynamics.

Our first paper on a related subject, \cite{Koziol2}, contains already some ideas about, and some evidence for, how the stress in MD simulations propagates into the sample volume. Since then, we concentrated on attempts to understand the physical mechanism and provide a mathematical description of pressure penetration. At the same time, we excelled in our technical skills, and improved the accuracy of simulations and data analysis. Seldom do we observe that the retarded nature of stress propagation is taken into account in simulations, therefore this aspect, the proper understanding of the simulated physical phenomena, is of the crucial importance. 

The harmonic theory used in the lattice vibrations of solids assumes that the anharmonic terms in the lattice potential energy
expansion are neglected, while the quadratic term is retained. This assumption has several consequences \cite{Ashcroft}, \cite{Kittel}, \cite{Katsnelson}, \cite{Cowley}: 
a) There is no thermal expansion of solids. b) Interaction between lattice waves does not exist, and a single elastic wave does not decay or change its form with time. c) Elastic constants do not depend on pressure and temperature. Hence, understanding the role of anharmonicity is the key factor for knowing the dynamic and thermodynamic properties of real materials. 

Non-linearity in the physics of crystals is omnipresent. Even small, non-linear contributions to inter-atomic potentials, may considerably influence the dynamics of particles, leading to chaotic motion and may cause seemingly unexpected energy losses \cite{Complexity}. It may cause localization of vibrations, depending on the strength of the non-linear contribution \cite{Herrera}. The first systematic studies of the role of these effects date back to over a half-century ago, with the methods of thermodynamic Green's functions \cite{Pathak}. Lattice dynamics and conditions of its stability have been addressed as well \cite{Lowell}. 

The presence of anharmonic potentials in materials is confirmed now by a broad spectrum of experimental techniques. X-ray diffraction and terahertz time-domain spectroscopy is used to resolve potential in molecular crystals \cite{Hutereau}, or higher harmonics detection in ultrasound propagation in steel \cite{Nucera} is used, or analysis of phonon oscillation spectra in graphene and other materials \cite{Hu}. Higher-order phonon non-linearities are observed by optical, infrared, and terahertz spectroscopy \cite{Hoegen}.

In MD simulations most of the potentials available may be treated as approximately only harmonic, in the best case. For instance, we found in our studies that Bonny \cite{Bonny} and Artur \cite{Artur} EAM potentials commonly used in simulations of steel have a small anharmonic contribution. In the case of Bonny's, an anharmonic addition may be approximated by $|x|^{2.5}$.

Many potentials are explicit anharmonic, like Morse potential \cite{Grado}, $E(r) \sim [\exp(-2\alpha(r-r_0))-2\exp(-\alpha(r-r_0))]$, Buckingham potential: $E(r) \sim [\exp(-r/\rho)-C/r^6]$ \cite{Carre}, or variants of Lennard-Jones potential \cite{Nucera}.

Hence, in this article, we study the simplest possible non-linear inter-atomic potentials of the form $V(x)= \epsilon_0 x^n$, where $n$>1, mostly it is between $n$=1.5 and $n$=3.60. Since closed-form analytical solutions to equations of motion in that case are not known in general, we perform simulations of the dynamics of particles by using an MD simulation tool, LAMMPS \cite{LAMMPS}. It is an overlooked opportunity that MD simulations may be used as a tool for inspiration and for guiding our understanding of physical processes, with the simulation results largely playing a role like that of experimental data. MD is often used in studies of lattice dynamics of solids \cite{Hellman}. 

In section 2 we describe the simulation setup and creation of non-linear potentials. 

Section 3 contains the classical derivation of a period of oscillations of a two-layer system for two simulation modes considered. Scaling relations are discussed between physical quantities such as displacement, velocity, and pressure. 

Section 4 presents MD results for larger structures with many crystallographic layers. The dynamics of virial stress profiles (waves of pressure) are discussed and compared to the results of a theoretical model of the chain of masses and springs available for harmonic crystals. The model was first solved mathematically by Erwin Schr\"{o}dinger in 1914 \cite{Erwin}, \cite{Muhlich}, and independently, by using another mathematical approach, by A. D. de Pater in 1974 \cite{Pater}. Both these fundamental works remain wildly unknown. Our simulation results provide the first evidence (to the best of our knowledge) for the validity of their solutions. The approach based on the model of masses connected by spring was found useful in case of explaining, for instance, Raman spectra in a few layers graphene \cite{Lui}, \cite{Tan}, and in our analysis of Van der Waals interaction in graphene \cite{Koziol}.

Finally, we discuss how the speed of sound is influenced by the non-linearity parameter $n$.

\section{Simulations setup, potentials, and samples.}
\label{Structure}

\begin{figure}[ht]
	\centering
	\includegraphics[scale=0.35]{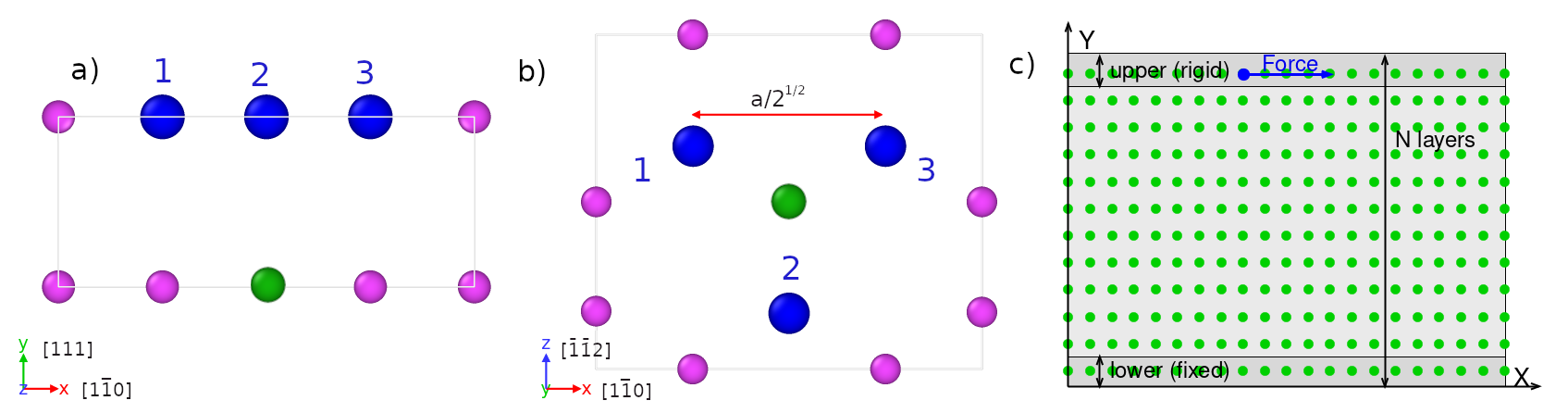}
	\caption{The orientation of FCC sample and distance between NN atoms. 
		a) shows mutual atoms' arrangement as seen on x-y plane and b) on x-z plane. c) shows the sample, with the same orientation as in figure a). The distance between green atom on the lower plane and 3 NN atoms on the next upper plane (blue atoms) is $a/\sqrt{2}$, and it is the same as the distance between blue atoms, shown by the red arrow in b). That distance is the translational period of potential in X-direction. 
	}
	\label{fig:neighbors00}
\end{figure}

Since we want to have the simulation conditions similar as much as possible to the situation we have already studied intensively and very accurately, we use a special configuration suitable when the dynamics of dislocations in FCC steel is studied.

We assume the FCC structure of the material, which is oriented in a specific way \cite{Koziol2}: Y direction is along [111] crystallographic axis, X is in [$1\bar{1}0$] and Z in [$\bar{1}\bar{1}2$]. In dynamical studies either an initial displacement of a layer of atoms is made in X direction, or pressure is applied to the upper sample layer in that direction, as shown by the label "Force" and arrow in \ref{fig:neighbors00} c). Figures \ref{fig:neighbors00} a) and b) show the position of atoms from two perspectives. 

In this orientation of the sample, it follows from symmetries that a possible contribution to forces in the Y direction equals 0 when a shear force is applied in the X-direction. 

In MD simulations we register (with desired time resolution, every few time steps) quantities like positions, velocities, pressure, stress, etc. Next, we compute the average values of these quantities for whole separate planes. Planes are enumerated from 1 at the top (upper region in Fig. \ref{fig:neighbors00}) to N for the lowermost layer (\texttt{lower} region). The number of atoms in every plane must be the same, and the uppermost region must contain one only layer, and also exactly the same number of atoms as all other layers. Otherwise, the simulation results obtained will be noisy and hard to interpretation. In particular, for instance, when the \texttt{upper} rigid region has a different than 1 number of layers, the time dependence of pressure on the sample surface as reported by LAMMPS will be entirely different, and it will depend also on the number of layers in that region. This fact is commonly miss-understood and ignored.

To achieve these conditions, the sample is checked for the same number of atoms in layers by using scripts that count atoms. Assigning atoms to layers is done by using their Y-coordinate. This method works well for samples with up to hundreds of layers when layer separation distance is properly chosen. Hence, finally, we must obtain a sample with the number of layers N and the total number of atoms in the sample must be an exact multiplicity of N. In the case of all samples studied here, every layer has 4800 atoms (samples have the size of 120 oriented unit cells in the X-direction and 20 in the Z-direction), while the number of layers, N, ranges from 2 to 33. The size of samples in X- and Z-direction does not play a role in our simulations (at least in cases when it is larger than a few unit cells and as long as the number of atoms in a single layer allows to achieve good statistical averages over physical quantities). Simulations performed in this work have been conducted at temperature T=0 (that allows us to avoid sample annealing). 

We use the \texttt{table pair style} in LAMMPS to provide interatomic potential. A text file is created for that by using Perl scripts. It contains the force and potential energy as a function of distance between particles. We aim, in this simple approximation, to account for forces between nearest neighbors (NN) only. Therefore the potential provided must be limited by an upper cut-off. We used 2.8 {\AA} for that. It means we are limited to some maximal values of the applied displacements or pressure. The cut-off on the lower side plays less role in simulations itself. We used 0.5 {\AA} for that (such a short cut-off is not needed, as we found out later). 

For the convenience of numerical computation, we use the following formula for the table-style potential used in LAMMPS:

\begin{equation}
	E_p(r)=\epsilon_0\left(-1+\left|(r-r_0)\right|^{n}\right),
\label{eq:potential}
\end{equation}

where $r$ and $r_0$ are in {\AA} and $n$ could be any positive value larger than 1. Most of the results presented here are for $n$ in the range 1.5-3.5. That potential is 0 at $r-r_0=1$ [{\AA}] and it has value $-\epsilon _0$ at $r=r_0$, for any $n$.

We aim to reproduce possibly closely the results for $n=2$ to these observed in steel in our previous work. Therefore, we choose $\epsilon_0=4.2$ eV at the minimum of the potential well, as in the case of the potential of Belland et al. \cite{Belland} characterized at the NIST\footnote{\href{https://www.ctcms.nist.gov/potentials/entry/2017--Beland-L-K-Tamm-A-Mu-S-et-al--Fe-Ni-Cr/2017--Beland-L-K--Fe-Ni-Cr--LAMMPS--ipr1.html}{NIST, Interatomic Potentials Repository}} repository. For particles, we assign the mass of one atom of Fe.

The arrangement of atoms is shown in \ref{fig:neighbors00} a) and b) (all atoms are in the equivalent surroundings), where crystal orientation in figure a) is the same as in c).

Let us concentrate on finding out the potential energy of the green atom.

If to assume that the position of the green atom is \textbf{$a_0$}=(0,0,0), 
we find that (in units of lattice parameter $a$) the positions of 3 NN (blue) atoms may be written as:
\textbf{$a_1$}=$(-1/(2\sqrt{2})$, $1/\sqrt{3}$,  $+1/(2\sqrt{2}\sqrt{3}))$, 
\textbf{$a_2$}=$(0, 1/\sqrt{3}$, $-1/(\sqrt{2}\sqrt{3})$ and \textbf{$a_3$}=$(+1/(2\sqrt{2})$, $1/\sqrt{3}$, +$1/(2\sqrt{2}\sqrt{3}$)).

When the green atom is moved in the X-direction for $x$, the distance between it and 3 blue atoms becomes:

\begin{equation}
	r_1(x) = a\cdot \sqrt{
	\left(\frac{-1}{2\sqrt{2}}+\frac{x}{a}\right)^2 + \left(\frac {1}{\sqrt{3}}\right)^2 + \left(\frac{1}{2\sqrt{2}\sqrt{3}}\right)^2}
	= r_0\cdot \sqrt{ 1 -\frac{x}{r_0} + \left(\frac{x}{r_0}\right)^2},
\label{eq:r1}
\end{equation}

\begin{equation}
	r_2(x) = a\cdot \sqrt{
	\left(\frac{x}{a}\right)^2 + \left(\frac {1}{\sqrt{3}}\right)^2 + \left(\frac{-1}{\sqrt{2}\sqrt{3}}\right)^2
	}
	=r_0\cdot \sqrt{
		1 + \left(\frac{x}{r_0}\right)^2}, 
\label{eq:r2}
\end{equation}

\begin{equation}
	r_3(x) = a\cdot \sqrt{
	\left(\frac{1}{2\sqrt{2}}+\frac{x}{a}\right)^2 + \left(\frac {1}{\sqrt{3}}\right)^2 + \left(\frac{1}{2\sqrt{2}\sqrt{3}}\right)^2
	} = r_0\cdot \sqrt{ 1 +\frac{x}{r_0} + \left(\frac{x}{r_0}\right)^2}, 
\label{eq:r3}
\end{equation}

where $r_0=a/\sqrt{2}$. 

Let us write: $r_i/r_0=\sqrt{1+y}$, where $i$ is 1,2 or 3. We have: $r_i^{'}(y)/r_0 = (1/2)\cdot (1+y)^{-3/2} \rvert _{y=0} = 1/2$. Therefore:

$(r_i(y) - r_0)/r_0 \approx \frac{1}{2}\cdot y$. Hence, to the lowest order of x:

\begin{equation}
	\frac{r_1(x) - r_0}{r_0} \approx -\frac{1}{2}\cdot \frac{x}{r_0}, ~~~\frac{r_2(x) - r_0}{r_0} \approx 0, ~~~
	\frac{r_3(x) - r_0}{r_0} \approx \frac{1}{2}\cdot \frac{x}{r_0}.
	\label{eq:r50}
\end{equation}

Since in 
\ref{eq:potential} $E_p$ is a function of the absolute value of $(r-r_0)$, by adding contributions from 3 atoms, we obtain as an approximation of $E_p(x)$, with respect to the potential minimum:

\begin{equation}
	E_p(x) -E_p(0) \approx 2\epsilon_0 \cdot \left|\frac{x}{2}\right|^n.
	\label{eq:potentialApprox}
\end{equation}

Let us notice that 
\ref{eq:potentialApprox} does not depend on $r_0$ in that lowest order approximation.

\begin{figure}[!ht]
	\centering
	\includegraphics[scale=1.0]{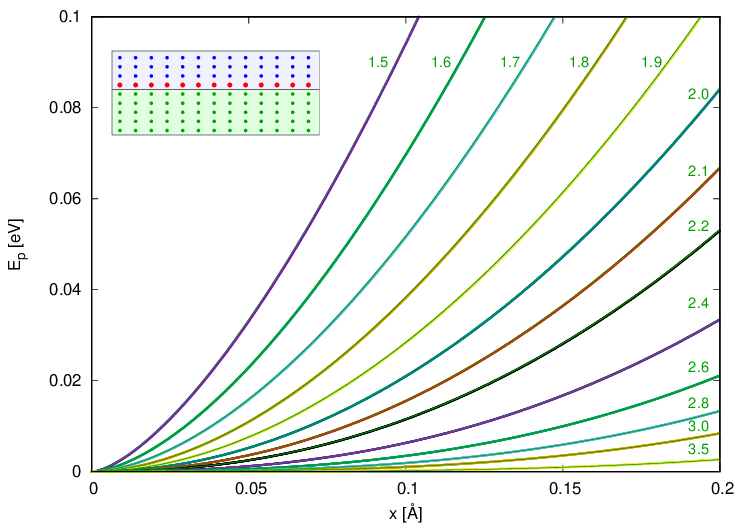}
	\caption{Potential energy change $E_p(x)$ of a single atom in a layer displaced for $x$ with respect to the other layer, for several values of the non-linearity parameter $n$, from $n$=1.5 to $n$=3.5, as shown by labels. On the top of sets of data points for any given value of $n$, tiny green lines are drawn by using $E_p(x)=2\epsilon_0 \cdot (x/\xi)^n$, with one pair of parameters $\epsilon_0=4.240$ eV, and $\xi$=2.006 {\AA}, common for all values of $n$.
		The inset 
		illustrates the method of computation of $E_p$ in LAMMPS: dots symbolize positions of
		atoms. There are three groups of atoms created: green, red and blue; each of them belonging to their own group. The blue group of atoms is
		moved in the X-direction for finding change in potential energy, which is computed for red atoms only.
	}\label{fig:Ep2_00}
\end{figure}

The potential has been checked in two ways. First, we calculated numerically $E_p(x)$ by using equations 
\ref{eq:potential}-\ref{eq:r3}, and we compared the results with explicit functions of type $x^n$. 

Next, we did that in LAMMPS, by determining changes in potential energy when the layers of atoms are displaced relative to each other. Both methods give the same results. In the range of $x$ between 0 and  0.2 {\AA}, and for n between 1.5 and 3.5, the same exponent is observed as designed (the least squares fit of potential determined from MD simulations to function  $x^n$ gives deviations in $n$ less than 0.15 \%), with the overall potential described by this kind of function (figure \ref{fig:Ep2_00}): 

\begin{equation}
	E_p(x)=2\epsilon_0 \cdot (x/\xi)^n, 
	\label{eq:potentialApprox2}
\end{equation}

where $\epsilon_0=4.240$ eV and $\xi$=2.006 {\AA}. When $n$ decreases below 1.5, the accuracy of fitting the data by using Eq. \ref{eq:potentialApprox2} deteriorates quickly. 

Hence, assuming inter-atomic potential between two NN atoms in the form of 
\ref{eq:potential} leads to the potential of interaction between any atom with all its NN on the crystallographic lattice in the form given by 
\ref{eq:potentialApprox} as an analytical approximation, or as 
\ref{eq:potentialApprox2}, as an approximation of MD simulation results.

The inset in Fig. \ref{fig:Ep2_00} illustrates the method of computation of $E_p$ in LAMMPS, as we used it: dots symbolize positions of
atoms. There are three groups of atoms created: green, red and blue; each of them belonging to their own group. The blue group of atoms is
moved in the X-direction for finding change in potential energy, which is computed for red atoms only. 

Attention must be paid to the proper determination of potential energy from MD simulations. 
Every atom has 12 NN, 6 of them on the same plane, and 3 on every of the two neighboring planes. Therefore, the total potential energy of an atom ought to be -12$\epsilon_0$. However, the potential energy per one atom, when $x=0$, as reported by LAMMPS, is exactly -25.2 eV, i.e., it is -12$\epsilon_0$/2 (where $\epsilon_0$=4.2 eV). When computing total potential energy of a system with a large number of atoms, and averaging it over all atoms, we obtain the same value, i.e., -12$\epsilon_0$/2. However, when we determine the potential energy change of a group of atoms in one layer, after it is moved (let say for red atoms in inset of figure \ref{fig:Ep2_00}), we ought to take into account that the total potential energy change of the entire system contains also contribution from the potential energy of green atoms as well, since these atoms are moved as well relative to red atoms. Change of their potential energy is exactly the same in value. It is easier to understand that by using the following example: let us assume that we have a system of two only atoms separated by distance $r_0$, i.e. both are at the minimum of their potential energy well. In that case their total potential energy is -$\epsilon_0$. However, LAMMPS will report their potential energy per atom as -$\epsilon_0$/2, a value which is two times lower. Hence, we conclude that the potential energy of a single atom that is used in any later calculations of their dynamics must be two times larger than the reported by LAMMPS potential energy per atom. Not taking that into account results in a disagreement by a factor $\sqrt{2}$ between the observed in simulations and the computed frequency of oscillation of atoms (layers).
The factor $\sqrt{2}$ is due to the scaling of oscillation frequency as square root of potential energy.
 
As a result of imposed conditions, the average X for the region \texttt{lower} does not change with time. Average Y for regions \texttt{lower} and \texttt{upper} remain the same.

\section{Two-layer system.}
\label{2layer}

\subsection{Period of oscillations.}
\label{Period}

The equation of motion for an anharmonic oscillator is nonlinear, making it more challenging to solve analytically compared to a simple harmonic oscillator.

One common approach to finding approximate solutions for an anharmonic oscillator is perturbation theory, where one starts with the solution for a related harmonic oscillator problem and then adds correction terms to account for the anharmonicity. 

It is important to note that the higher the value of $n$, and the higher the order of the perturbation theory considered, the more complex the calculations become. In some cases, the perturbation series may not converge well, leading to difficulties in obtaining accurate results.

Another method is to use numerical techniques, such as solving the differential equation of motion numerically using computer algorithms like the iterative Runge-Kutta method (see, e.g., \cite{Butcher}).

Finding exact, general solutions for potentials of that form, and approximate ones, have been discussed by Euler \cite{Euler}, Amore \cite{Amore}, Harko \cite{Harko}. They use the idea of performing a non-linear transformation of variables $x$ and $t$. The general solution of a class of equations of that type can be expressed as an infinite sum of components with the hypergeometric function \cite{Harko}. That however lacks the simplicity to be analyzed.
Approximate solutions are found for a logarithmic form of anharmonicity, which is shown to be related to conventional power-law anharmonicity $|x|^n$ in the limit $n\rightarrow 0$, \cite{Znojil}.

Variants of the chain of masses and springs model have been used often to study aspects of non-linearities, chaos, and localization/de-localization of lattice vibrations  \cite{Nucera}, \cite{Barreto}, \cite{Pankov}, \cite{Kashchenko}. The model gained a broad interest after the Fermi-Pasta-Ulam paradox was published in 1955 \cite{FPU}. 

There is another way of gaining an understanding of solutions to the problem, and it is based on the use of molecular dynamics simulations. The method is flexible, and allows one to approach problems more and more accurately by introducing or removing some physical interactions and/or constraints, approximating them gradually to the real. 

Before going into the analysis of the results of MD simulations, let us concentrate on presenting the known analytic solution to the problem of simple anharmonic oscillator with the power-law dependence of potential energy on the displacement between two particles.

We are interested in the following potentials:

\begin{equation}
	V(x) = \epsilon_0 \cdot \left| \frac{x}{\xi}\right|^{n},
	\label{eq:T_V3}
\end{equation}

where $\xi$ is a certain length scale (actually, already determined in the previous section). The absolute value of displacement, $\left|x\right|$, is there only for mathematical rigor. As we will see, in our case we will not need to integrate in a negative x-direction.

From the equation of motion, $m\ddot{x}=-dV(x)/dx$, we obtain an integral of the motion $E$, which is the total energy:
\begin{equation}
E=\frac{m\dot{x}^{2}}{2}+V(x)  \label{eq:energy}
\end{equation}

It is well known \cite{Amore}, \cite{Landau} that the time of the motion may be given as an integral over the trajectory of motion $C$ in the $x-v$ phase space:

\begin{equation}
t=\int_C dt= \int_C \frac{dt}{dx} dx = \int_C \frac{dx}{v(x)}.
\label{eq:t_v}
\end{equation}

In our case, the periodic motion of the particle is restricted to the
interval $x_{-}<x<x_{+}$, where the turning points $x_{\pm }$ satisfy $
V(x_{\pm })=E$; that is to say, $\dot{x}=0$ at those points, i.e. the kinetic energy becomes zero and $E$ converts to the potential energy.
Moreover, for any potential $V(x)$ that is symmetric with respect to its minimum value at $x=0$, 
the motion during one period consists of traversing the closed trajectory, for instance: $x_{+}\rightarrow 0$, $0\rightarrow x_{-}$, $x_{-}\rightarrow 0$, and $0\rightarrow x_{+}$. Equations of motion for any of these 4 trajectories are the same. Therefore, the period is 4 times the time of the movement of the particle between $x=0$ and $x=x_{+}$. Hence, since from 
\ref{eq:energy} $v^2=(2/m)(E-V(x))$, after substituting $v(x)$ into \ref{eq:t_v}, we have:

\begin{equation}
T=4 \cdot \sqrt{\frac{m}{2E}}\int_{0}^{x_{+}}\frac{dx}{\sqrt{1-V(x)/E}}
\label{eq:T_V}
\end{equation}

We can introduce a dimensionless variable $z$ representing the ratio of the potential energy to the total energy:

\begin{equation}
z=V(x)/E = \frac{\epsilon_0}{E} \cdot \left(\frac{x}{\xi}\right)^{n}, ~~~
dx= \left(\frac{E}{\epsilon_0}\right)^{1/n} \cdot \frac{\xi dz}{n z^{(n-1)/n}}  . 
\label{eq:T_V4}
\end{equation}

Now, at the turning point $x_+$ potential energy $V(x_{+})$ equals to $E$, hence $z=1$ in 
\ref{eq:T_V4} at that point. Therefore the 
 integration range in 
\ref{eq:T_V} changes from [$0$:$x_{+}$] to [0:1] when $z$ is used instead of $x$ as an integration variable. We can write 
\ref{eq:T_V} in the form:

\begin{equation}
T=4\cdot \sqrt{\frac{m}{2E}} \cdot  \left(\frac{E}{\epsilon_0}\right)^{1/n} \cdot \frac{\xi}{n}  \int_{0}^{1}     \frac{dz}{ z^{(n-1)/n}     \sqrt{1-z}}.
\label{eq:T_V6}
\end{equation}

By introducing parameters $p=1/n$ and $q=1/2$, we have:

\begin{equation}
T=4\cdot \sqrt{\frac{m}{2E}} \cdot  \left(\frac{E}{\epsilon_0}\right)^{1/n} \cdot \frac{\xi}{n}  \int_{0}^{+1} z^{p-1} (1-z)^{q-1} dz.
\label{eq:Z}
\end{equation}

We recognize that the integral expression in Eq. 
\ref{eq:Z} is the definition of Euler's $\beta $ function \cite{Abramowitz}: 

\begin{equation}
\beta(p,q) = \int _0 ^1 x^{p-1} (1-x)^{q-1} dx = \frac {\Gamma(p) \Gamma(q)}{\Gamma(p+q)}, ~~~p>0, q>0.
\label{eq:G}
\end{equation}

Hence, we have:

\begin{equation}
T=4 \cdot \xi \sqrt{\frac{m}{2\epsilon_0}} \cdot  \left(\frac{E}{\epsilon _0}\right)^{\frac{2-n}{2n}} \cdot \Psi(n), ~~~  \Psi(n) = \frac{1}{n} \cdot \frac {\Gamma(1/n) \Gamma(1/2)}{\Gamma(1/n+1/2)}.
\label{eq:Z2}
\end{equation}

The introduced $\Psi(n)$ is a slowly decreasing function of $n$, with $\Psi(1)=2$, $\Psi(2)=\pi/2$.

At the maximum displacement (at $x=A$, amplitude of oscillations), based on 
\ref{eq:T_V3}, the potential energy equals the total energy $E$, and we get $A/\xi =\left(E/\epsilon_0\right)^{1/n}$. Therefore, we can relate the period of oscillations $T$ to the amplitude $A$:

\begin{equation}
T=4 \cdot \xi \sqrt{\frac{m}{2\epsilon_0}} \cdot (A/\xi)^{(2-n)/2} \cdot \Psi(n).
\label{eq:Amp}
\end{equation}

Accordingly, the angular frequency can be given as:

\begin{equation}
\omega = \frac{\pi}{2\xi} \cdot \sqrt{\frac{2\epsilon_0}{m}} \cdot \frac{1}{\Psi(n)} \cdot (A/\xi)^{(n-2)/2}.
\label{eq:Omega}
\end{equation}

When $n>0$, there are three cases known to us when close-form exact analytic solutions are available: 1) A trivial case when $n=1$, resulting in a force not dependent on the distance between particles, describing the motion of mass objects on an inclined plane. 2) An ideal harmonic oscillator for $n=2$. 3) For $n=4$ a closed-form solution can be expressed with Jacobi elliptic functions \cite{Lawden}, \cite{Whittaker}.

\subsection{Displacement method.}
\label{Dmethod}

\begin{figure}[!ht]
	\centering
	\includegraphics[scale=1.0]{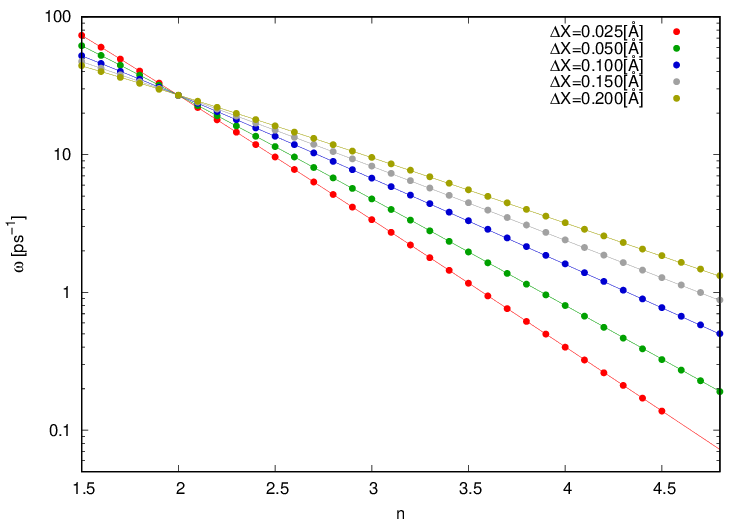}
	\caption{Dependence of oscillation frequency $\omega$ on the exponent of potential, for displacements from 0.025 to 0.2 {\AA}, as shown in the legend. 
		The fitting lines are functions based on Eq. 
		\ref{eq:Omega}, $\omega(n) = a/\rm{\Psi(n)}\cdot (\Delta X/b)^{(n-2)/2}$, where parameters $a$ and $b$ are common for all curves. 
	}\label{fig:Disp01}
\end{figure}

Two basic methods may be used for studying the nonlinear properties of anharmonic crystals. First, we would like to know well the dynamics of interaction between two layers, only. For that, we use the displacement method: a structure as shown in \ref{fig:neighbors00} c) is used with only two layers, the upper and the lower. The upper layer is moved in X-direction for a certain distance $\Delta X$ (much less than the periodicity of the crystal), away from the equilibrium position, and at time $t=0$, it is released free while all atoms are undergoing integration in NVE statistical ensemble. The upper layer is treated as a rigid body, while the lower layer is fixed. This is a realization of the idea of a classical oscillator where a particle of mass $m$ enters a periodic in-time motion in a potential well. The quantities determined in simulations are the coordinates of all atoms, their velocities, potential energy, components of virial stress tensor. We perform averaging of those quantities for all atoms in every layer. It ought to be noticed that the default resolution used in LAMMPS for reporting quantities such as potential energy, positions, and velocities of atoms ought to be increased for 2-3 orders to achieve the accuracy of simulations as reported by us.

\subsubsection{Frequency of oscillations.}

The period of oscillations $T$ may be determined from the dependence on time of the surface pressure, $P_{xy}(t)$, from the displacement of the upper layer $\Delta X(t)$, or its velocity $v_x(t)$. Usually, we use the least-squares fitting method to find out the period by approximating the simulation data with the function $\sin(2\pi t/T)$. Albeit the fitted data in general, deviate strongly from that function, this is an accurate method for finding the oscillation period.

The scaling of the oscillation frequency on the exponent of potential is shown in \ref{fig:Disp01}, for a few values of the initial displacement, by using 
\ref{eq:Omega} in the form:
$\omega(n,\Delta X) = a/\rm{\Psi(n)}\cdot (\Delta X/b)^{(n-2)/2}$. Here, $a=\pi/2\xi \cdot \sqrt{2\epsilon_0/m}$, and $b=\xi$.
Notice that the parameter $\epsilon_0$ in Eq. \ref{eq:Omega} must be multiplied by 2 when computing frequency in agreement with the potential experienced during 
MD simulations, as expressed by Eq. \ref{eq:potentialApprox2}.
The least squares fitting of all the data gives values of $a$=42.35 ps$^{-1}$ and $\xi$=2.020 {\AA}.
The value of $\xi$ is in agreement with what is found from potential analysis $E_p(x,n)$ in the previous section. 
When $\epsilon_0$=$2\cdot 4.24$ eV and $\xi$=2.006 {\AA} are used, $a=(\pi/2\xi) \cdot \sqrt{2\epsilon_0/m}$=42.4 ps$^{-1}$ is obtained.

\subsubsection{Fourier analysis.}

\begin{figure}[!ht]
	\centering
	\includegraphics[scale=1.0]{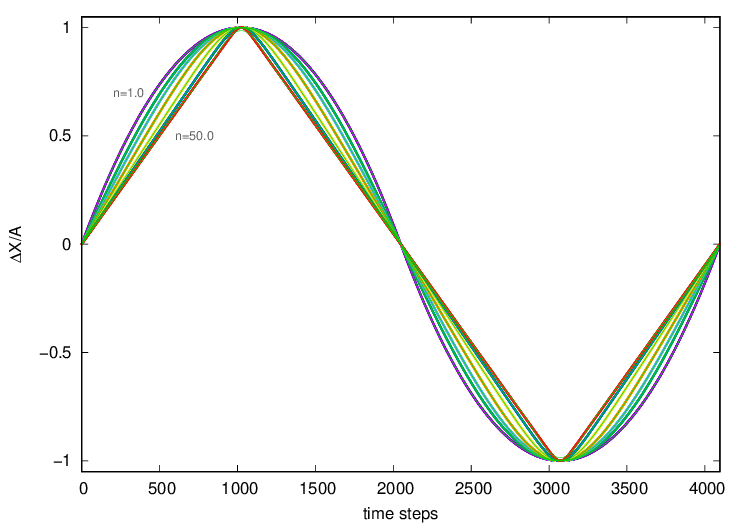}
	\caption{Comparison between computed displacement curves (normalized by the amplitude of oscillations $A$), shown by different colours, and the data reconstructed from Fourier components (up to 13th harmonic, all odd), shown by tiny green lines. The results are shown for the following values of $n$: 1.0, 2.0, 3.0, 5.0, 10.0, 20.0, and 50.0. The period of oscillations is $2^{12}$=4096 time steps. When $n$=1, $X(t)$ can be constructed exactly by parabolic curves. When $n\rightarrow \infty$, $\Delta X(t)/A$ tends to a triangle-like function.
	}\label{fig:DZ8192_01}
\end{figure}

Displacement curves as a function of time step are shown in Fig. \ref{fig:DZ8192_01}. These data were prepared to find out
how the components of a Fourier series describing $X(t)$ evolve when $n$ changes. We used fast Fourier transform (FFT) method. It requires a sequence of data points that are equally spaced in time and their number must be a power of 2, therefore we have $2^{12}$=4196 points for one period of oscillations. 

In the case of the displacement method, we performed FFT analysis of two kinds of data sets of $X(t)$ sequences. First, we constructed $X(t)$ dependencies numerically. Let us explain how that can be done.

When integration in \ref{eq:Z} is done to $z<1$, the resulting function is called the incomplete beta function:

\begin{equation}
	\beta_z(p,q)= \int_{0}^{z} x^{p-1} (1-x)^{q-1} dx, 
	~~~ 0 \le x \le 1,
	\label{eq:ZInc}
\end{equation}

It allows us to find time as a function of $z$, $t(z)$, (where $z(x)$ is given by \ref{eq:T_V4}). We obviously would like to have the reverse relation, $x(t)$. For that, the inverse incomplete beta function is used, usually named as $I(z,p,q)$ when it is regularized. We compute it numerically, for instance with \texttt{betaincinv} from Python module \texttt{scipy.special}. One ought to be aware that this function computes the 1/4th of the trajectory (similarely as $\beta_z$ in Eq. \ref{eq:ZInc} gives us $t(x)$ for 1/4th of $T$). Therefore, it is necessary to join properly together results of four similar calculations for obtaining the full oscillation period, as shown in Fig. \ref{fig:DZ8192_01} for a few values of $n$. 

The second type of data sets analyzed by FFT are those obtained from MD simulations. MD results are practically identical to those obtained by the numerical method. However, obtaining reliable and accurate simulation data requires attention to certain details. First of all, we need to find out an exact value of the oscillation period $T$. In the next step, we ought to adjust the time step in LAMMPS script so that it equals exactly $T/4196$, in our case. 

For FFT computations, we used Perl module \texttt{Math::FFT}. Instead, one can use for instance Python libraries through \texttt{scipy.fft}.

In our case, it is actually not necessary to use FFT for finding Fourier series amplitudes. An alternative method, giving a comparable accuracy, is to use \texttt{fit} command available in Gnuplot. One can approximate $X(t)$ dependence by a finite sum of harmonics of $\sin(t)$ and/or $\cos(t)$, let say in the form: $X(t)/A= \sum_{i=0}^{k}a_{i}\cdot \sin(2\pi \cdot i \cdot t/T)$. The \texttt{fit} command can find out coefficients $a_{i}$ for $k$ equal up to a few tens.

\begin{figure}[!ht]
	\centering
	\includegraphics[scale=1.0]{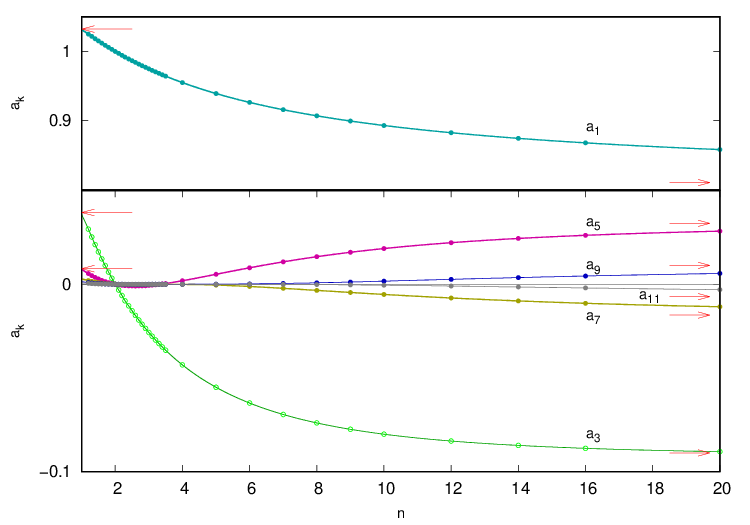}
	\caption{Fourier coefficients $a_k$ for periodic functions of $\Delta X(t)/A$ as shown in Fig. \ref{fig:DZ8192_01}, 
		with $\Delta X(t)/A=\sum_{i=0}^{\infty}a_{2i+1}\cdot \sin(2\pi(2i+1)t/T)$. All even coefficients are equal to zero. 
		Arrows on the left indicate values of $a_k$ for $n=1$ when $\Delta X(t)/A$ becomes parabolic-like. Arrows on the right
		show values of $a_k$ in the limit of infinite $n$, when $\Delta X(t)/A$ becomes triangle-like function.
		The lines are for guiding eyes, only.
	}\label{fig:DZ8192_10}
\end{figure}

Since functions shown in Fig. \ref{fig:DZ8192_01} are asymmetric with respect to $t=0$, the Fourier decomposition must contain only $\sin(t)$ type of components. 
Moreover, due to the property $X(t)=-X(T-t)$, only odd components will be present. Figure \ref{fig:DZ8192_10} shows the dependencies of amplitudes $a_k$ on the non-linearity parameter $n$ when the overall functions are defined as $\Delta X(t)/A=\sum_{i=0}^{\infty}a_{2i+1}\cdot \sin(2\pi(2i+1)t/T)$.

When $n<2$, values of all $a_k$ are positive, while for $n>2$, there are values of both signs. Our data suggest that some $a_k(n)$ change signs more than once (as may be seen in this scale for $a_5(n)$ only). In the limit when $n=1$, $X(t)$ can be constructed from parabolic dependencies. It is easy to verify by computing proper Fourier integrals that in this case, all odd coefficients are given by a series $a_k = 32/\pi^2 \cdot 1/k^3$ (see also \cite{Folland} and \cite{Herman}). Red arrows on the left in Fig. \ref{fig:DZ8192_10} show these values for $k=1,3,5$. In the limit of $n\rightarrow \infty$, $\Delta X(t)/A$ tends to a triangle function of time. In that case, we found by integration that the Fourier series amplitudes are given as $a_k=8/\pi^2 \cdot (-1)^{(k-1)/2}/k^2$. The arrows on the right show these limiting values.

It ought to be noticed that Fourier series amplitudes $a_k$ do not depend on the amplitude of oscillation $A$, when we normalize them by $A$, and time is normalized by $T$.

\subsubsection{Relationships between displacement, velocity and pressure.}

Since $z$ in Eq. 
\ref{eq:ZInc} is defined as $(\epsilon_0/E) \cdot (x/\xi)^n$, we can find out the dependence $t(x)$. In particular, at $z \ll 1$ we can use an approximation: $\beta_z(p,q) \approxeq (1/p)z^p(1-z)^q \approxeq (1/p) z^p$. Therefore, based on Eq. 
\ref{eq:Z}, we can write for $t \ll T/4$ (near the potential minimum):

\begin{figure}[!ht]
	\centering
	\includegraphics[scale=1.0]{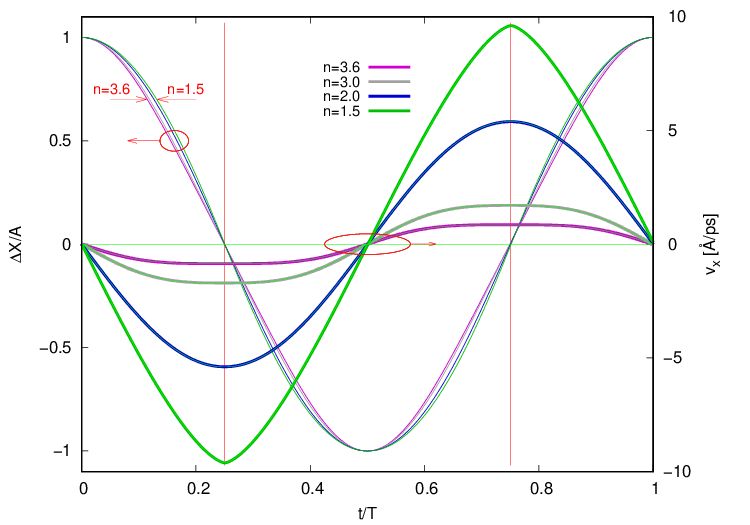}
	\caption{Displacement normalized by its initial value (amplitude, 0.2 {\AA} in this case) is shown as a function of time normalized by T. Additionally, velocities $v_x$ as a function of normalized time (right vertical axis), are shown. Maxima or minima of $v_x$(t) are found exactly at t=T/4 or at t=3T/4, as indicated by vertical red lines. On the top of $v_x$(t) curves denoted in the legend, we draw thin green lines $v_x$(t) computed from $\Delta$X(t) data by using the formula: $v_x(t)= v_{max}(n) \cdot \sqrt{1-\left|\Delta X(t)/A\right|^n}$ (taking care of proper sign, depending on time range).
	}\label{fig:DZ00A}
\end{figure}

\begin{equation}
t(x) \approxeq \sqrt{\frac{m}{2E}} \cdot  \left(\frac{E}{\epsilon_0}\right)^{1/n} \cdot \frac{\xi}{n}
\cdot n \cdot z^{1/n} = \sqrt{\frac{m}{2E}} \cdot x.
\label{eq:TXApprox}
\end{equation}

Using that equation, we obtain velocity at potential minimum, which is the maximal (in value), $v_{max} = \sqrt{2E/m}$. It does not depend on $n$ (as desired): it depends on the total energy $E$. Since $E={\epsilon_0}\cdot (A/\xi)^n$, we have a scaling relation: 

\begin{equation}
v_{max}=\sqrt{2\epsilon_0/m} \cdot (A/\xi)^{n/2}.
\label{eq:vScaling}
\end{equation}

That dependence is valid for data as included in \ref{fig:DZ00A}, and it can be approximated (for amplitude $A$ ranging from 0.005 {\AA} to 0.2 {\AA}) as $v_{max}(n) = 54.4\cdot (A/\xi)^{n/2}$ {\AA}/ps, with $\xi$=2.02 {\AA}. Notice that the proportionality coefficient, 54.4{\AA}/ps differs from that in $\omega$ scaling in \ref{fig:Disp01} by a factor of 1.286 {\AA}, which is exactly $2\xi/\pi$, from the scaling factor in  
\ref{eq:Omega}.

\begin{figure}[!ht]
	\centering
	\includegraphics[scale=1.0]{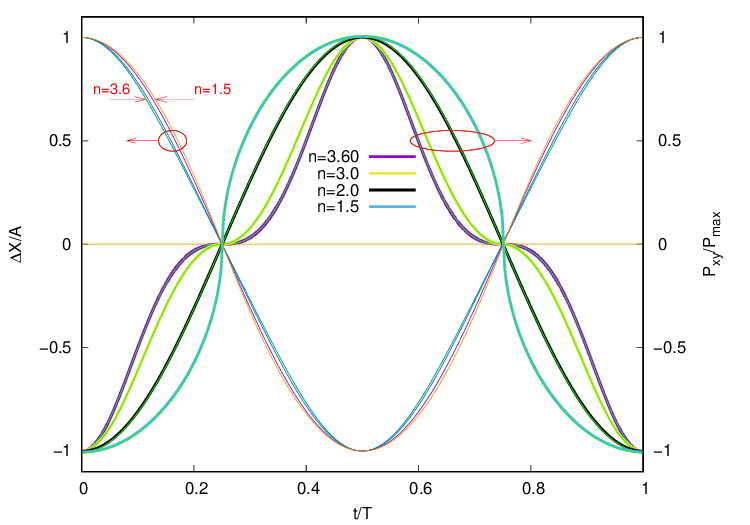}
	\caption{Relationship between pressure (force) and  displacement. Displacement raised to power (n-1) reproduces pressure curves (with a certain rescaling factor). The response has been recorded as an initial displacement of the upper layer for 0.2 {\AA} for a single period of oscillations, which is normalized to 1 for all curves. Results are shown for 4 values of n: 1.5, 2.0, 3.0, and 3.6. Three kinds of curves are drawn: displacement of the upper layer $\Delta X /A$ (left vertical axis), pressure at the surface  $P_{xy}$ normalized by its maximum value (right vertical axis), and displacement raised to power $n-1$, $(\Delta X/A) ^{n-1}$, normalized in such a way that the curves coincide with pressure values. $(\Delta X/A) ^{n-1}$ curves are drawn on the top of pressure curves and largely cover them up. The colours in the legend refer to pressure data.
	}\label{fig:D00}
\end{figure}

Figure \ref{fig:DZ00A} shows a relationship between velocity and displacement. Displacement is normalized by its amplitude (left vertical axis), and velocity (right vertical axis) is not. Both quantities are shown as a function of $t/T$, for 4 values of $n$. Vertical red lines are located at $t=T/4$ and $t=3T/4$, intersecting $\Delta X(t)$ at $\Delta X$=0 and $v_x(t)$ at its minimum and maximum values. Velocity depends on kinetic energy, $v_x=\sqrt{2E_k/m}$, where $E_k=E-V(x)$ and $V(x)=\epsilon_0 (x/\xi)^n$. From that, the relation follows: $v_x(t)= v_{max}(n) \cdot \sqrt{1-\left|\Delta X(t)/A\right|^n}$. It is drawn with thin green lines on the top of $v_x$ data from MD simulations.

Similarly, we show a relationship between displacement and the surface pressure $P_{xy}$. In LAMMPS pressure components are computed as a sum of forces (averaged over the surface area) exerted on the surface from within the entire sample volume. In our particular case, there are two only layers of atoms. $P_{xy}$ is registered for the upper layer and it is caused by the forces acting on it from the lower layer; see  \ref{fig:neighbors00} c). This illustration helps us understand the physical significance of pressure in LAMMPS. Hence, while $V(x)=\epsilon_0 (x/\xi)^n$, force is given by the derivative of potential, $F=dV/dx=n\epsilon_0/\xi \cdot (x/\xi)^{n-1}$. Therefore, displacement raised to a power ($n-1$) reproduces pressure curves, with a rescaling factor that is related in this case to $n\epsilon_0/\xi$. In \ref{fig:D00}, the response has been recorded to an initial displacement of the upper layer for 0.2 {\AA}, for a single period of oscillations. Results are shown for 4 values of $n$: 1.5, 2.0, 3.0, and 3.6. Three kinds of curves are drawn: displacement of the upper layer $\Delta X$ normalized by its amplitude (0.2 {\AA} in this case; left vertical axis), pressure at the surface  $P_{xy}$ normalized by its maximum value (right vertical axis), and displacement to power $n-1$, $(\Delta X/A) ^{n-1}$, normalized in such a way that the curves coincide with pressure values. Computed $(\Delta X/A) ^{n-1}$ curves are drawn on the top of pressure curves, by using tiny green lines that largely cover them up.

\subsection{Heaviside method.}
\label{Hmethod}

\subsubsection{Potential curves and frequency of oscillations.}

\begin{figure}[!ht]
	\centering
	\includegraphics[scale=1.0]{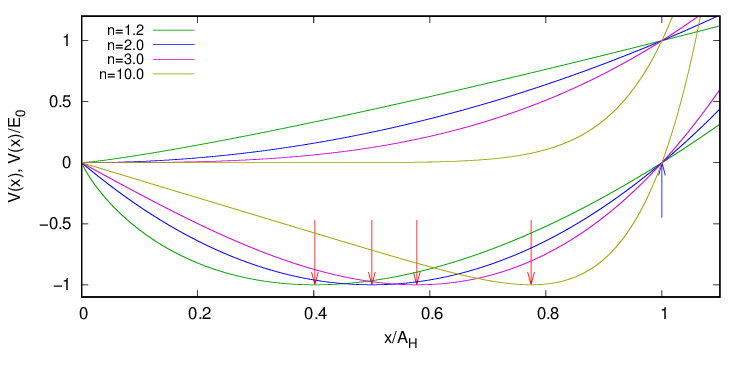}
	\caption{Potential energy curves without the force applied (the upper 4 curves) and with the force of F=1 applied (the bottom 4 ones), for n=1.2, 2,0, 3.0, and 10.0, when $\epsilon_0=1$ and $\xi=1$, according Eq. \ref{eq:VF}. The blue arrow indicates maximum displacement (amplitude $A_H$), and red arrows show positions $x_0/A_H$, i.e. minima of potential energy when force is applied. 
	}\label{fig:P07}
\end{figure}

The method described in the previous section, with two layers, where one of them is initially displaced, is suitable for determining the fundamental oscillation frequency related to the potential between layers, as we did for graphene \cite{Koziol} and steel \cite{Koziol2}. However, in typical MD simulations, we rather investigate the response of material to an applied pressure. The results in that case are, in general, different from those derived to this point. They would be the same only in the case of a perfect harmonic potential. If any non-linearities are present, the effective potential of interaction between layers after applying pressure becomes non-symmetric concerning the minimum of the potential well (\ref{fig:P07}). 

Hence, when pressure (force $F$) is applied in a Heaviside-type dependence on time, to a sample, as shown schematically in \ref{fig:neighbors00} c), the upper layer is moved away from the equilibrium position at $x=0$ and tends to achieve another equilibrium position at a position $x_0$ in  \ref{fig:P07}. Oscillations occur  around $x_0$. That equilibrium would be reached after a (possibly very long) simulation time when losses were present. In the simplest model we are studying we find no indication for the presence of any energy losses at temperature $T=0$ and therefore the equilibrium $x_0$ is not supposed to be reached ever. In simulations, it is easy however to introduce non-physical losses during oscillation that allow us to determine quickly the positions of equilibrium, $x_0$ \footnote{A convenient way of introducing artificial energy losses is by adding NVT integration alongside NVE one. An attempt to stabilize temperature (possibly low) will cause the loose of kinetic energy of atoms. The efficiency of this method can be controlled by adjusting \texttt{timestep}, temperature and other parameters related to the NVT \texttt{fix} command.}. 

In the presence of a Heaviside type of force $F$, since force is a derivative of potential, we may modify potential given by 
\ref{eq:T_V3} to the following form:

\begin{equation}
V(x) = \epsilon_0 \cdot (x/\xi)^{n} -Fx.
\label{eq:VF}
\end{equation}

Let us assume that the total energy of the particle is zero at $x=0$ and $x=A_H$ (see \ref{fig:P07}). We find the amplitude of oscillations $A_H$ from the condition $V(A_H)=0$, and $x_0$ is determined from the condition $dV/dx=0$, resulting in:

\begin{equation}
	\frac{A_H}{\xi}=\left(\frac{\xi F}{\epsilon_0}\right)^{\frac{1}{n-1}}, ~~~ \frac{x_0}{\xi}=\left(\frac{\xi F}{n\epsilon_0}\right)^{\frac{1}{n-1}}, ~~~\frac{x_0}{A_H}=\left(\frac{1}{n}\right)^{\frac{1}{n-1}}.
\label{eq:AMP}
\end{equation}

Notice that $x_0/A_H$ tends to 0 when $n \searrow 0$, and it tends (very slowly) to 1 when $n \nearrow \infty$, with $x_0/A_H=1/e \approx0.3679 $ for $n=1$. Also, the ratio $x_0/A_H$ does not depend on the applied force $F$. It depends on $n$, only. 

The depth of the potential well is $E_0 = V(x_0)$:

\begin{equation}
	E_0 = \epsilon_0 \cdot (x_0/\xi)^{n} -Fx_0 = (1-n) \epsilon_0 \cdot  \left( \frac {\xi F}{n\epsilon_0}\right) ^{\frac{n}{n-1}} = (1-n) \epsilon_0 \left(\frac{x_0}{\xi}\right)^{n}.
\label{eq:E0}
\end{equation}

We may assume now that the total energy of the particle is $E=|E_0|$. At the bottom of the potential well (at $x=x_0$), $E$ converts into fully kinetic energy, while at $x=0$ and $x=A_H$ the maxima of potential energy are reached during particle oscillation.

Let us notice that for $n \rightarrow \infty$, $V(x)$ is dominated by the $Fx$ term when $x \lesssim x_0$, i.e. the potential there is well approximated by a linear one, as illustrated in \ref{fig:P07} by the curve for $n$=10.

This time we are not able to provide exact expressions for the period of oscillations since we do not know how to perform integration analogous to that one in \ref{eq:T_V} for potential given by Eq. \ref{eq:VF}. However, we expect that some scaling properties derived so far will remain valid for the new potential. Hence, let us have approximate expressions for $\omega$ and $T$. By substituting $|E_0|$ instead of $E$ in  
\ref{eq:Z2}, we have a new period of oscillations:

\begin{equation}
T=4 \xi \sqrt{\frac{m}{2\epsilon_0}} \cdot  
(n-1)^{\frac{2-n}{2n}} \cdot \left( \frac{\xi F}{n\epsilon_0}\right)^ {\frac{2-n}{2(n-1)}}
\cdot \Psi(n),
\label{eq:TF}
\end{equation}

and the angular frequency:

\begin{equation}
\omega= \frac{\pi}{2\xi \cdot \Psi(n)} \cdot \sqrt{\frac{2\epsilon_0}{m}} \cdot
(n-1)^{\frac{n-2}{2n}} \cdot \left( \frac{\xi F}{n\epsilon_0}\right)^ {\frac{n-2}{2(n-1)}}.
\label{eq:OmegaF}
\end{equation}

\subsubsection{Phase space portraits.}

\begin{figure}[!ht]
	\centering
	\includegraphics[scale=1.0]{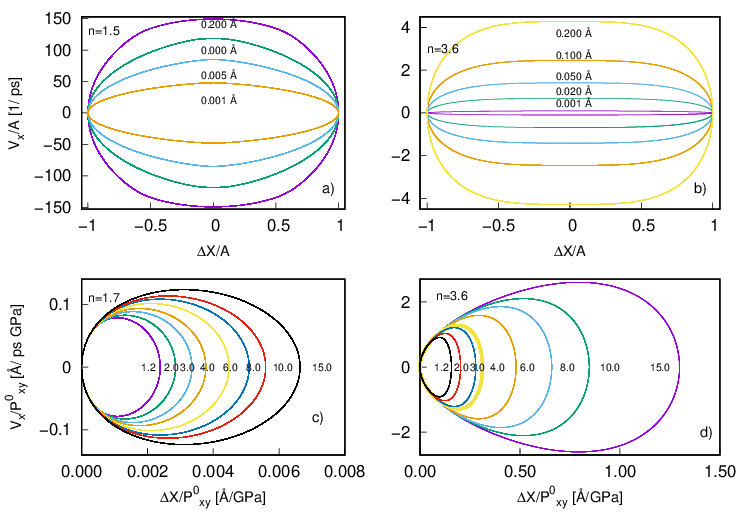}
	\caption{Phase space portraits of the system for different situations. The dependence of velocity on the displacement is shown. The upper two figures a) and b) are for the displacement method of simulations, with amplitudes of displacement as shown in the legend. Both, the displacements and velocities are normalized by amplitude $A$.
		The c) and d) parts are when Heaviside pressure is applied, for pressure values as shown in the legend (in GPa). The displacements and velocities are normalized by the value of applied pressure $P^0_{xy}$.
	}\label{fig:ZCirclesXX}
\end{figure}

A good way of comparing the dynamics in the case of both methods (displacement, and Heaviside) is by drawing phase space portraits, as in  \ref{fig:ZCirclesXX}. This is also a convenient way of testing the validity of ongoing simulations. The diagrams show the dependencies of velocity versus displacement. The upper two figures are for the displacement method, where we see phase loops symmetric around both X and Y with the center at (0,0). For the Heaviside method, the loops are symmetric only concerning Y axis. 

In MD simulations, atoms movement for regions \texttt{upper} and \texttt{lower} must be restricted to X-direction only. Without these restrictions, minimal  asymmetries in the sample structure or fluctuation effects due to temperature will drive the motion into quasi-chaotic oscillations (still preserving the attractor points). Such a quasi-chaotic motion results in sub-harmonic components in any time dependencies, that is, in $\Delta X(t)$, $v_x(t)$, $P_{xy}(t)$, etc. That would manifest itself in  phase space diagrams as unclosed loops changing with time and filling in densely a large portion of phase space.

\subsubsection{Fourier series analysis.}

\begin{figure}[!ht]
	\centering
	\includegraphics[scale=1.0]{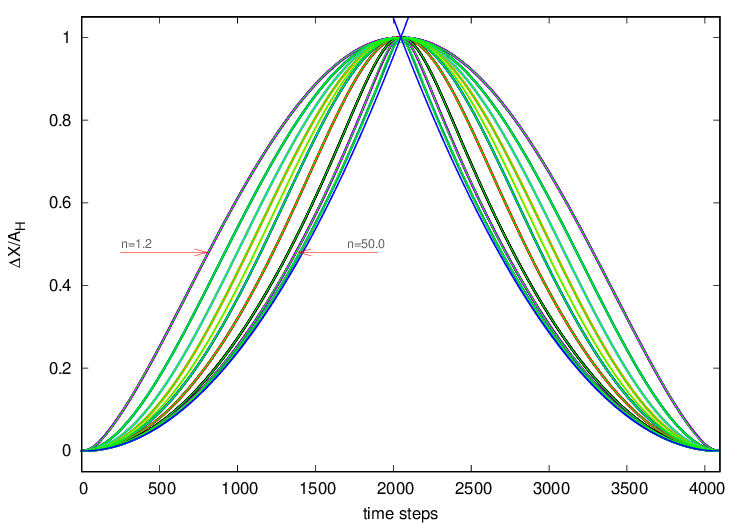}
	\caption{Displacement curves (normalized by the amplitude of oscillations $A_H$) in the Heaviside method, obtained by using MD simulations are shown by different colours. Tiny green lines represent the data obtained by reconstruction from Fourier components (up to 11th harmonic), all odd). The results are shown for the following values of $n$: 1.2, 1.5, 2.0, 2.5, 3.0, 5.0, 10.0, 20.0, and 50.0. The period of oscillations is 4096 time steps. 
		When $n\rightarrow \infty$, $X(t)$ tends to a parabolic function, shown by blue curves.
	}\label{fig:PZ8192_01}
\end{figure}

Preparing quality MD simulation data for analysis of Fourier harmonics by using Heaviside method requires care in the proper choosing of the applied pressure value. At small values of $n$, close to 1, the amplitude of oscillations is very small and a large pressure must be applied and short \texttt{timestep} used. At large values of $n$, the situation is reversed. The displacement curves shown in Fig. \ref{fig:PZ8192_01} have been obtained for pressures in the range $10^{-4}-10^{10}$ Pa and for \texttt{timestep} parameter between $10^{-5}$ and $10^{10}$ ps. The dependencies $\Delta X(t)$ are however not sensitive to the value of applied pressure when time is normalized by $T$ and the displacement by $A_H$. Let us pay attention that the normalization by $A_H$ differs from the normalization by the amplitude $A$ used in the case of the displacement method. Now the computed Fourier series amplitudes must by multiplied by 2 when comparing the FFT data between both methods.

\begin{figure}[!ht]
	\centering
	\includegraphics[scale=1.0]{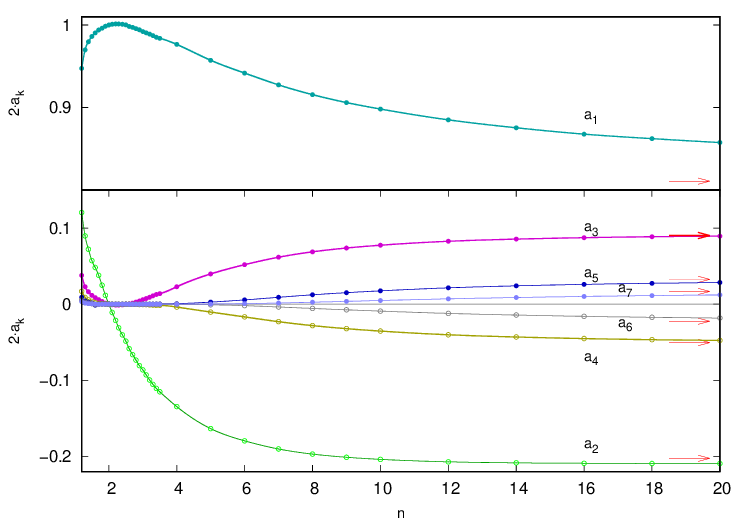}
	\caption{Fourier coefficients $a_k$ for periodic functions of $\Delta X(t)/A_H$ as shown in Fig. \ref{fig:PZ8192_01}, 
		with $\Delta X(t)/A_H=-\sum_{k=0}^{\infty}a_{k}\cdot \cos(2\pi\cdot k \cdot t/T)$. 
		When $n<2$, values of all $a_k$ are positive, while for $n>2$ the odd and even harmonics have the opposite sign.
		When $n\rightarrow \infty$, $X(t)$ tends to a parabolic function, as shown by the two crossing blue curves. 
		The calculated Fourier integrals in that limiting case give us: $a_k = -(8/\pi^2) \cdot (-1)^k /k^2$. Arrows on the right are drawn at levels corresponding to values of these coefficients.
		The lines are for guiding eyes, only.
	}\label{fig:PZ8192_10}
\end{figure}

Due to the symmetries of $X(t)$ we conclude right away that only harmonics of $\cos(t)$ will contribute to the displacement function. Moreover, since $X(t)=X(T-t)$, even harmonics must be present.

When $n\rightarrow \infty$, $X(t)$ tends to a parabolic function, as shown by the two crossing blue curves in Fig. \ref{fig:PZ8192_01}. The reason for that is that the potential well for $x<x_0$ may be approximated by a linear dependence on the displacement, as illustrated by the curve for $n=10$ in Fig. \ref{fig:P07}, and $x_0\rightarrow A_H$ at the same time.  This allows us to determine what are the limiting values of amplitudes of harmonics for large values of $n$. We performed proper calculation of Fourier integrals in that case, obtaining that $\Delta X(t)/A_H$ can be represented by the Fourier series of coefficients $a_k = -(8/\pi^2) \cdot (-1)^k /k^2$ (see also \cite{Folland} and \cite{Herman}). Arrows on the right in Fig. \ref{fig:PZ8192_10} are drawn at levels corresponding to values of these coefficients.

In the case of $n\searrow 1$, when $n$ becomes close to 1,
MD simulation results become unreliable, for reasons that are not fully understood by us, and therefore we show results for $n$ starting from 1.2, and higher. 

As it is seen in Fig. \ref{fig:PZ8192_10}, for $n<2$ all Fourier components are of the same sign, while for $n>2$ the odd and even harmonics have the opposite sign.

\subsubsection{Relationships between the displacement, velocity and pressure.}

\begin{figure}[!ht]
	\centering
	\includegraphics[scale=1.0]{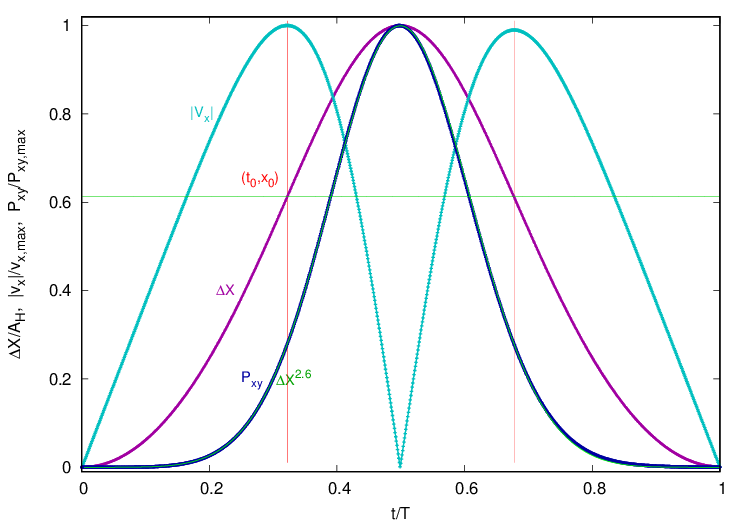}
	\caption{The applied pressure is 500 MPa (that value is not significant in this case). Displacement, the absolute value of velocity, and the surface pressure are shown, when $n$=3.6. All quantities are normalized by the period of oscillations $T$, and by their maximal values. Red vertical lines cross velocity curves at their maximum values. The green horizontal line is drawn at $x_0/A_H$. On the top of pressure data, the computed 
	$(\Delta X(t)/A_H)^{n-1}$ is drawn with a tiny green line. 
	}\label{fig:P00}
\end{figure}

Asymmetry of potential well manifests strongly in the shape of all time-dependencies of the Heaviside method. 
Figure \ref{fig:P00} shows how pressure reflects the force acting on the upper layer that comes from the potential derivative.

The red vertical lines intersect velocity curves at their maximum values. The green horizontal line is drawn at $\Delta X = x_0/A_H$ computed with 
\ref{eq:AMP}. The point of crossing of red and green lines, labelled as $(t_0,x_0)$, is the coordinate of the potential minimum. Velocity there has an extreme value. $t_0$ is also the inflection point on displacement and pressure curves. When $n>2$, as in this case, $t_0/T$ is larger than 1/4 and $x_0/A_H$ is larger than 1/2.

The pressure registered at the upper layer, $P_{xy}$, is the result of forces acting on that layer from the lower layer. Instead of using the mass of the entire layer and its surface, it is convenient in any calculations to use the average mass of a single atom and its averaged surface area per atom, $S$. Hence, $P_{xy}$ related to the force (a derivative of potential), may be written as:

\begin{equation}
P_{xy}(t)= -(n\epsilon_0/\xi)\cdot (x(t)/\xi)^{n-1} \cdot (1/S). 
\label{eq:XAH00}
\end{equation}

The maximum pressure at the surface is when $x=A_H$ is reached, and it is given by: 

\begin{equation}
P_{xy,max}=-(n\epsilon_0/\xi S)\cdot (A_H/\xi)^{n-1}.
\label{eq:XAH0}
\end{equation}

Therefore, we can write immediately:

\begin{equation}
\frac{P_{xy}(t)}{P_{xy,max}} = \left(\frac{\Delta X(t)}{A_H} \right)^{n-1}.
\label{eq:XAH}
\end{equation}

Hence, normalized displacement raised to a power $(n-1)$ reproduces pressure curves normalized by the maximum pressure value, as is shown by tiny green lines drawn over the data points of pressure in \ref{fig:P00}. At the point $(t_0,x_0)$ velocity reaches the maximum (or minimum) value. While we know values of $x_0$ from \ref{eq:AMP}, we do not know values of $t_0$, since we have no exact solutions of equations of motion.

We can draw some additional conclusions regarding the pressure curves. 

At the point $(x_0,t_0)$ we have $P_{xy}(x_0) = (n\epsilon_0/\xi S) \cdot  (x_0/\xi)^{n-1}$. Therefore $P_{xy}(x_0)/P_{xy,max} = \left( x_0/A_H\right)^{n-1}$. Since $x_0/A_H=(1/n)^{1/(n-1)}$ (Eq. 
\ref{eq:AMP}), we can write:

\begin{equation}
\frac{P_{xy}(x_0)}{P_{xy,max}} = \frac{1}{n}.
\label{eq:XAH2}
\end{equation}

That relation can be verified by the data in \ref{fig:P00}. Note only that green lines are drawn with the left vertical axis, while pressure is described by the right axis. Therefore, one should use 
the vertical red lines indicating $t_0$ to determine the crossing point of $x_0$ by pressure curves.

Moreover, $x_0$ is the equilibrium position that would have been reached when losses of energy existed. Hence, pressure at that point, $P_{xy}(x_0)$ is compensated by the applied pressure $P^0_{xy}(x_0)$.
Therefore, the ratio of the maximal value of pressure oscillations $P_{xy,max}$ to the value of the surface pressure applied $P^0_{xy}$, equals $n$, as we observe indeed in simulations:

\begin{equation}
P_{xy}(x_0) = P^0_{xy},~~~\frac{P_{xy,max}}{P^0_{xy}} = n.
\label{eq:XAH3}
\end{equation}

Surface area per atom is $S=a^2 \sqrt{3}/4=5.49\cdot 10^{-20}m^2$, where $a$ is the lattice constant, $a$=3.56 {\AA}, The proportionality coefficient between pressure and displacement in Eq. 
\ref{eq:XAH00}, $\epsilon_0/(S\xi)$, is 122.4 GPa when computed with $\epsilon_0$=8.48 eV. This value is consistent with what we find  from the analysis  of dependence between $P_{xy,max}$, and $A_H$ for various pressures and $n$'s.

\subsubsection{Scaling of oscillation frequency with n and pressure.}

\begin{figure}[!ht]
	\centering
	\includegraphics[scale=1.0]{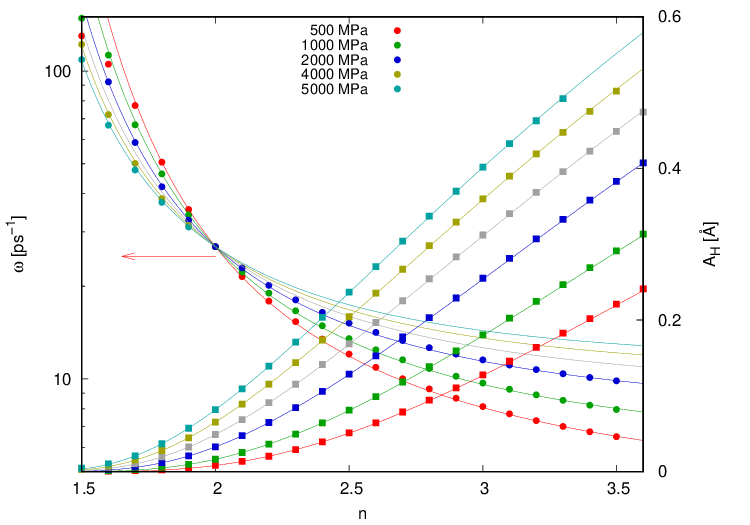}
	\caption{Scaling of oscillation frequency $\omega$ with the exponent of potential when simulations are performed under constant pressure, for pressure values as shown in the legend. 
		The left vertical log scale is for $\omega(n)$ (full circles) and the right linear scale is for amplitude $A_H(n)$ (squares). The colours of symbols and lines are the same for $\omega(n)$ and for $A_H(n)$. 
	}\label{fig:N2alpha01}
\end{figure}

Figure \ref{fig:N2alpha01} shows the scaling of oscillation frequency and amplitude for simulations performed under constant pressure, for a broad range of pressure values. 

Based on Eq. \ref{eq:OmegaF}, we may write: $\omega(n,P)=a/\Psi(n) \cdot (n-1)^{\frac{n-2}{2n}} \cdot (bP/n)^ {\frac{n-2}{2(n-1)}}$, where $P$ is pressure in GPa. Here, the parameter $a$ has the same meaning as used in the decription of data in figure \ref{fig:Disp01}, $a=\pi/2\xi \cdot \sqrt{2\epsilon_0 /m}$. 
By using least squares fitting to all of the data, we obtain $a$=42.25 ps$^{-1}$, while in the case of figure \ref{fig:Disp01} that value was 42.4 ps$^{-1}$. The difference is insignificant and it is within the range of fitting error. For $b$ we find $b$=0.0195 GPa$^{-1}$. 

$A_H(n,P)$ in figure \ref{fig:N2alpha01} is fitted to all the data, also with two only parameters that are the same for all curves. Based on  
Eq. \ref{eq:AMP}, we may write: $A_H(n)= \xi \cdot \left(bP\right)^{\frac{1}{n-1}}$, where $b$ has the same meaning as in expression on  $\omega(n,P)$: $b=\xi S/\epsilon_0$.

The lines for $A_H(n,P)$ are drawn with $\xi$=1.97 {\AA} and $b$=0.0083 GPa$^{-1}$. Here, $\xi$ is very close to the value found already for a symmetric potential, of about 2.02 {\AA}. The value of $b$ is about 2.35 times smaller than the one from fitting $\omega(n,P)$. However, this value is consistent with what is obtained from computing $b=\xi S/\epsilon_0=0.0082 GP^{-1}$, with $\xi$=2.02 {\AA} and $\epsilon_0$ = 8.48 eV. It agrees also with the inverse parameter, 122.4 GPa, determined already from $P(A_H)$.

While equation \ref{eq:AMP} on amplitude scaling must be correct, the one on frequency ought to be treated as an approximation only, as mentioned already when deriving equations \ref{eq:TF} and \ref{eq:OmegaF}.

In summary, well-established scaling relations are valid in the case of the Heaviside method. If we change the force $F$ (pressure $P^0_{xy}$) by a factor $\lambda$, $F \rightarrow \lambda F$, the new potential depth can be written as $E_{0,\lambda} = \lambda^{\frac{n}{n-1}}\cdot E_0$. When simultaneous change is made to scales of distance and time, equations of motion remain the same for new quantities
(Eqs. \ref{eq:AMP}-\ref{eq:TF}):

\begin{equation}
	\begin{split}
		x_{0,\lambda} = \lambda^{\frac{1}{n-1}}\cdot x_0,~~~
		A_{H,\lambda} = \lambda^{\frac{1}{n-1}}\cdot A_H, \\
		T_{\lambda} = \lambda ^{\frac{2-n}{2(n-1)}} \cdot T,~~~
		\omega_{\lambda} = \lambda ^{\frac{n-2}{2(n-1)}} \cdot \omega.\\
	\end{split}
	\label{fig:Scaling}
\end{equation}

For instance, either by analyzing the ratio of $x$ to $t$ or by using the energy rescaling we find that velocities are transformed in this way: $v_{\lambda}\rightarrow \lambda ^{\frac{n}{2(n-1)}} \cdot v$. Acceleration will scale up as a square of velocity: $a_{\lambda}\rightarrow \lambda ^{\frac{n}{n-1}} \cdot a$. Since it is related to the force (pressure) acting on the particle, the same scaling will be valid for pressure as well: $P_{\lambda,xy}\rightarrow \lambda ^{\frac{n}{n-1}} \cdot P_{xy}$.

\clearpage

\section{Anharmonic crystals.}
\label{Ncrystals}

The description of the two-layer motion of crystallographic layers is the introduction only to the study of the dynamics of many-layer structures and will serve us as theoretical background for finding out how stress (sound) propagates in crystals. The model we use is essentially a 1-dimensional model of a chain of masses connected by springs, with masses representing crystallographic layers. That model has been found to describe well the stress penetration as analyzed by MD simulations in steel, it helped improve the accuracy of data analysis from simulations, as well it was helpful in developing a methodology for the simulation itself. 

The first ideas about the usefulness of this model were proposed in \cite{Koziol2}, where also some evidence can be found for the existence of pressure oscillations within the sample interior. Details of this model and supporting data analysis will be published separately.

\subsection{Dependencies on the number of layers N.}
\label{N-dependencies}

So far, we introduced the period of oscillations (and angular frequency) for two interacting layers. However, in the case of multi-layer materials, we have yet another period of oscillation and angular frequency. Once the pressure is applied to the upper surface of the sample, as shown in \ref{fig:neighbors00}c), it causes oscillations of the upper layer. That one interacts with the second one, etc. In other words, a pressure wave is induced at the surface, and that wave propagates into the material interior, with the speed of sound (in our case, it is the transverse sound mode). Once the wave reaches the lower layer of the sample, it becomes reflected there (depending on the boundary conditions imposed on that lower layer). The reflected wave interferes with the incoming wave and reaches the upper sample surface. The process repeats. In cases when no losses are present, a perfect periodic dependence on time is found for quantities such as the displacement of layers, their velocity, pressure at the surface, etc. That period of oscillations depends, however, on the size of the sample in the direction of wave propagation (Y-direction in our case). As a measure of sample size, we use the number of layers N and the period related to sound waves traversing the sample will be denoted as $T_N$. 

\begin{figure}[!ht]
	\centering
	\includegraphics[scale=1.0]{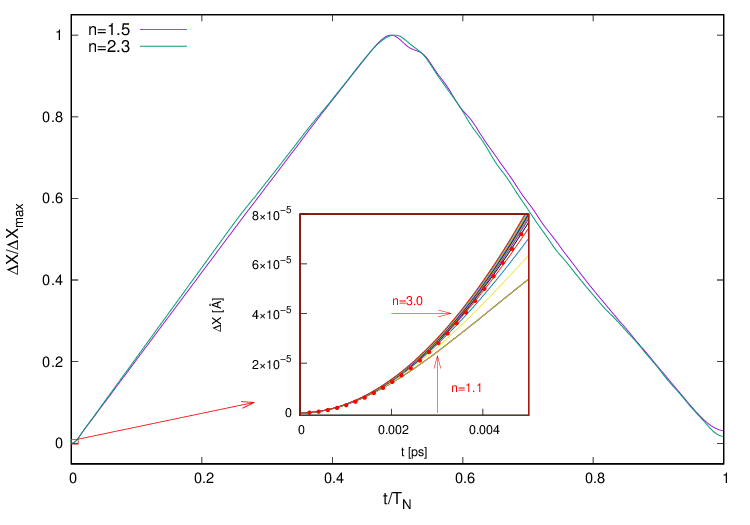}
	\caption{Displacement of the upper layer normalized by its maximum value as a function of time normalized by the period of oscillations $T_N$, for n=1.5 and n=2.3, at 1000 MPa, for a sample with N=33 layers. Nearly triangle-like dependencies of $\Delta X(t)$ are observed for any values of $n$. The inset shows expanded displacement (not normalized by $\Delta X_{max}$) as a function of time (not normalized by $T_N$) for a large number of curves for different $n$, for very short times. Red dots there have been computed for $n$=2 by using  
\ref{eq:un}.
	}\label{fig:aHeavy_150}
\end{figure}

We will show some results obtained for samples of changing sizes, from N=2 to N=20, with most simulations obtained for a sample with N=33.

Typical displacement versus time for one period of oscillations, where both quantities are normalized, $\Delta X(t)$ by its maximum value $\Delta X_{max}$ and time by the period of oscillations $T_N$, is shown in figure \ref{fig:aHeavy_150}. Pressure at the surface, $P_{xy}(t)$, follows closely $\Delta X(t)$ dependence.

The inset shows expanded displacement as a function of time (not normalized by $T_N$) for very short times. In that short time region, $\Delta X(t) \sim t^2$ is found for any $n$. Moreover, $\Delta X$ scales up linearly with the value of applied pressure, in the same time range. For $n$=1.1 this approximation works well below around 0.002 ps. For most curves shown it is valid below around 0.01 ps or more, and for all curves it is valid for $t/T$ less than around 0.1, where $T$ is the oscillation period for the 2-layer system. Hence, in that region layer dynamics may be viewed as being caused by constant force. This is the same as observed in figure \ref{fig:P00} for $t \ll T$. The red dots in the inset of \ref{fig:aHeavy_150} have been calculated for $n$=2, by using 
\ref{eq:un}, as will be explained.

\begin{figure}[!ht]
	\centering
	\includegraphics[scale=1.0]{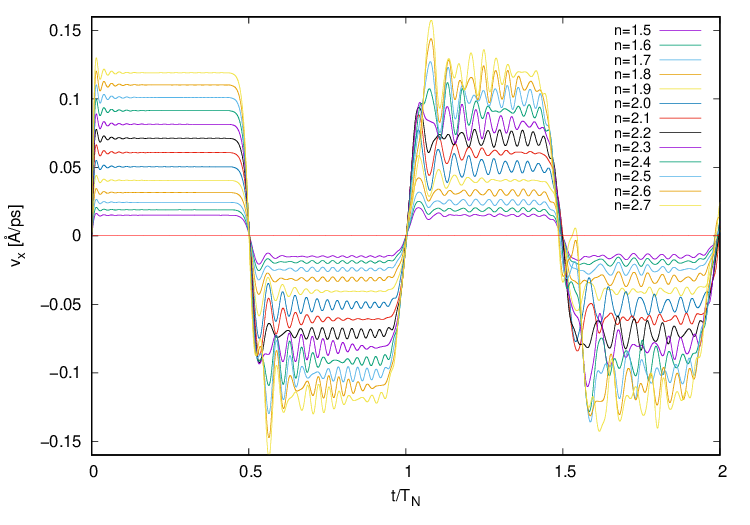}
	\caption{The velocity of the uppermost layer at 200 MPa pressure applied, for a sample with N=33 layers, when time is rescaled by the period of oscillation, for the first 2 oscillation periods.
	}\label{fig:aHeavyZ02M}
\end{figure}

\begin{figure}[!ht]
	\centering
	\includegraphics[scale=1.0]{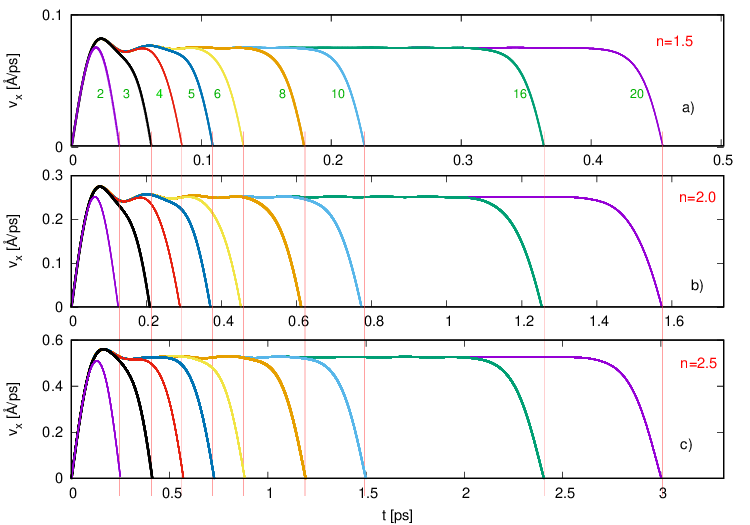}
	\caption{Velocity $v_x$ determined from simulations under pressure 1000 MPa is shown for times $t<T_N/2$ for three values of n: 1.5, 2.0, and 2.5, for figures from top to bottom. There are 9 curves drawn for different numbers of layers, from N=2 to N=20, as shown with green labels in the top figure. 
	}\label{fig:aHeavyCUT00}
\end{figure}

The velocity of the uppermost layer is shown in \ref{fig:aHeavyZ02M} for two periods of oscillations. We see that while the amplitude of velocity changes strongly with $n$, the time dependence $v_x(t)$ remains approximately the same for all curves when drawn with time normalized by $T_N$. 

The one full period of oscillations takes the time needed for 4 times of traversing by the wave the sample length in Y direction. At $t=T_N/4$ and $t=3T_N/4$ we usually notice a slight change in curves, when reflection of wave occurs from the \texttt{lower} layer, especially when $n$ is larger than 2 and the applied pressure is high. 

The points when $v_x(t)$ passes through a value of 0 can be used for an accurate determination of the half of the period of oscillations, $T_N/2$ (\ref{fig:aHeavyZ02M}). We have verified that $T_N$ determined in this way depends linearly on N, $T_N=\alpha \cdot (N-1)$, where $\alpha$ depends on the parameter of potential non-linearity $n$, i.e. the sound wave propagates similarly as in case of harmonic potential, when $n$=2. Period of oscillations (and speed of sound) depend on the applied pressure, though, when $n \ne 2$.

In \ref{fig:aHeavyCUT00} velocity of the uppermost layer is shown for the first half of the oscillation period for samples with different number of layers, as labeled in part a), ranging from N=2 to N=20. The curves for larger N cover up the data of smaller samples. We compare the results for 3 values of n. The range of the horizontal axis, in every case of different $n$, has been chosen in such a way that the vertical red lines cross $v_x$=0 axis at $T_N/2$ for all values of $n$. This illustrates that the period of oscillations, $T_N$, scales up in the same way with the number N of layers in samples, regardless of the non-linearity parameter $n$.  
Let us notice that for N=2 the curves $v_x(t)$ look quite the same, regardless of $n$. For N>2 the curves are nearly identical as well (the difference is in their amplitude). This suggests that the observed oscillations and their similarity is not related to the non-linearity parameter $n$.

\subsection{Pater's model of the chain of masses and springs for n=2.0.}
\label{Pater}

The analytical model of a chain of masses and springs that correctly describes our results for $n$=2, is provided in a little-known, old article by Pater \cite{Pater}. We will use equations from that paper in the analysis of our simulation results. However, the model of a chain of masses and springs dates back as far as the beginning of the XX century, to these times when the differences between continuous and discrete mechanics were formulated, with the first exact analytical results attributed to Erwin Schr\"{o}dinger \cite{Erwin},\cite{Muhlich}. The ingenious work of Schr\"{o}dinger remained however not known as well. 

Pater's exact analytical results reproduce properly the expected dynamics in the limit of continuous medium. In that case, for instance, the wave entering an elastic material exposed to surface Heaviside-type pressure continues to penetrate the bulk of the material, preserving its original shape. In the case, however, of a discrete medium, as the situation of our interests, the solutions are given in terms of Bessel functions. Let in our case $m$ enumerate particles in the chain, starting from number 1 as that one exposed to the initial Heaviside force. In our situation, crystallographic layers play the role of these particles. The quantities like displacements $u_m$, velocities $v_m$, and force $f_m$, acting on a particle $m$, is given by:

\begin{equation}
	u_m = \frac{F_1}{\rm{m}\Omega^2}\left[J_{2m}(2\theta) + \sum _{k=2m+2,2m+4,...} ^{\infty} (k-2m+1) J_k(2\theta)\right],
	\label{eq:un}
\end{equation}

\begin{equation}
	v_m = \frac{F_1}{\rm{m}\Omega}\left[J_{2m-1}(2\theta) + 2\cdot \sum _{k=2m+1,2m+3,...} ^{\infty} J_k(2\theta)\right],
	\label{eq:vn}
\end{equation}

\begin{equation}
	f_m = -F_1\cdot \left[ 1-J_0(2\theta) -2\cdot \sum _{k=2,4,...} ^{2m} J_k(2\theta) + J_{2m}(2\theta)\right],
	\label{eq:fn}
\end{equation}

where $J_k$ are Bessel functions of the first kind, and $\theta = \Omega t$.  The $\Omega$ there, for $n=2$, is $\sqrt{2\epsilon _0/\rm{m}}$, $F_1$ is the force applied at the first layer, the surface one, and \rm{m} is the mass of the layer. 

These equations are valid for $t<T_N/4$, i.e., before the front of the wave reaches the opposite side of the sample, where reflection occurs. The equations could still be used for longer times, however, we want to avoid complicating this description. 

In MD simulations, pressure at the surface (as reported by LAMMPS), $P_{xy}$, is a volume average of the internal virial stress, $\langle S_{xy}\rangle/V$, where $V$ in this definition is an average volume occupied by an atom. Virial stress  can be treated as a summation of forces exerted on the surface by all layers inside the sample volume. Some authors consider virial stress as a not well-defined quantity, with no clear physical meaning. It is, however, a statistical quantity, like temperature, valid as well for non-equilibrium systems, as long as these can be assumed as ergodic. The literature contains a reach discussion of the subject  \cite{Yang}, \cite{Zhou}, \cite{Subramaniyan}, \cite{Zimmerman}, \cite{Elder}. 

We have checked accurately and verified that $S_{xy}$ may be used consistently in MD simulations. In particular, for instance, the relation $P_{xy}=\langle S_{xy}\rangle/V$ provides a valuable test of the validity of simulations: any discontinuities or change of slope in that dependence strongly indicates problems with the simulations. Finding the proportionality coefficient between these quantities is the best way to determine the average volume occupied by one atom. 
\footnote{Virial stress components, $S_{ij}$, are reported by LAMMPS in units of bars, which are pressure units. However, $S_{ij}$ is computed as a product of pressure and volume $V$, and only after dividing $S_{ij}$ by $V$ we obtain the result in pressure units. The typical value of $V$ in the case of steel with an FCC structure is around 11 {\AA}$^3$.}

\begin{figure}[!ht]
	\centering
	\includegraphics[scale=1.0]{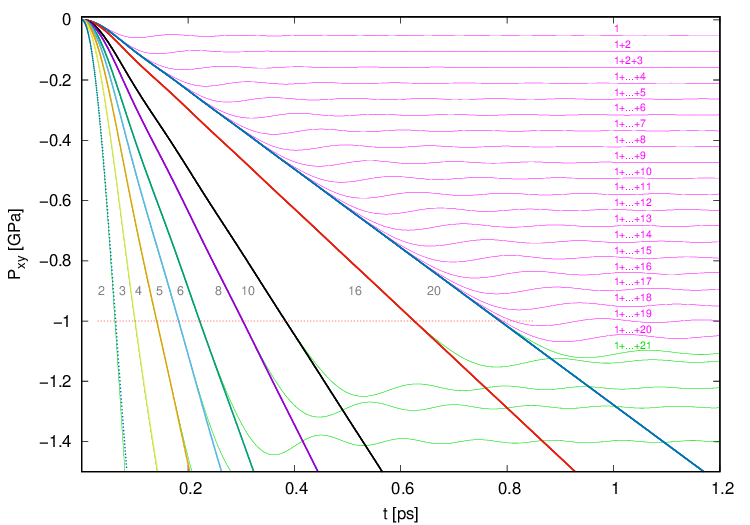}
	\caption{Pressure computed with Pater equations is compared with MD simulation results of $P_{xy}(t)$, under 1 GPa Heaviside pressure impulse, applied  to samples with N layers, where N is between 2 and 20 (as shown by horizontally aligned labels), when non-linearity parameter is $n$=2.00. The MD results are drawn with (densely packed) dots over green and magenta curves representing results obtained with Pater equations.
		Green curves show surface pressure based on 
\ref{eq:sumfnPROPER}, computed for N=2...20. 
		The horizontal red broken line at pressure -1 GPa is crossed by lines when $t=T_N/4$, and the numerical results from Pater equations (green lines) are supposed to be valid only for the data lying above that line.
		Magenta lines are computed for the sample with N=20 layers. The labels (vertically aligned) denote the number of layers over which the summation of the pressure contribution was taken into account. 
	}
	\label{fig:PressureScaling00}
\end{figure}

Hence, we identify $f_m$ as given by 
\ref{eq:fn} to be related to the virial stress averaged over all atoms  of layer $m$, $\langle S_{xy,m}\rangle $. Therefore, we can write:

\begin{equation}
	P_{xy}(t) = \frac{1}{\rm{S}}\sum_{k=1,2,...} ^{m} f_k(t) 
	= \sum_{k=1,2,...} ^{m} \langle S_{xy,k}\rangle/V,
	\label{eq:sumfn}
\end{equation}

where \rm{S} is the surface area of the uppermost layer, and the summation is valid for times $t < T_N/4$, i.e., for times less than the time needed for the front of the pressure wave to reach the opposite surface of the sample. The maximal allowed range of index $m$ in the above summation is N-1, since there are a maximum of N-1 layers contributing to the pressure at the surface layer \footnote{We could continue with $m$ approaching 2N, until $t<T_N/2$, but we want to avoid complicating the description by considering the reflection of waves from surfaces.}. In a case when the equilibrium of forces (stresses) was reached in the sample volume, the surface pressure would equilibrate N-1 contributions from the pressure of N-1 interior layers. That is why we ought to include a weight 1/(N-1) in the above summation. Therefore, the proper form of \ref{eq:sumfn} for computing the pressure contribution of layers ought to be:

\begin{equation}
	\begin{split}
P_{xy}(t) &= \frac{1}{N-1} \sum_{k=1,2,...} ^{N-1} \langle S_{xy,k}\rangle/V \\
 &=	-\frac{P^0_{xy}}{N-1} \sum_{k=1,2,...} ^{N-1} \cdot \left[ 1-J_0(2\theta) -2\cdot \sum _{i=2,4,...} ^{2k} J_i(2\theta) + J_{2k}(2\theta)\right],
	\end{split}
\label{eq:sumfnPROPER}
\end{equation}

where we substituted Eq. 
\ref{eq:fn} as an expression on forces from $k$-th layer and used there $P^0_{xy}$ instead of $F_1/\rm{S}$, where $P^0_{xy}$ is the amplitude of the applied surface Heaviside pressure.

Computational results of $P_{xy}(t)$ by using this method are shown in \ref{fig:PressureScaling00}, where comparison with MD simulations are included, for samples with the number of layers from N=2 to N=20.
That figure illustrates two aspects of computing surface pressure by using  
\ref{eq:sumfnPROPER}. First, with green lines we illustrate how the slope of $P_{xy}(t)$ changes with N. It is governed by the 1/(N-1) factor in  
\ref{eq:sumfnPROPER}, only (neglecting the obvious dependence on the amplitude of applied pressure, $P^0_{xy}$, and the timescale of the horizontal axis, which is determined solely by $\Omega$). The second aspect of calculations is illustrated by magenta lines, for the sample with N=20 layers.
 We use successive approximations of $P_{xy}(t)$ based on \ref{eq:sumfn}, by summing a finite number of contributions from forces $f_m(t)$, where $m$ denotes the number of layers (counting from the surface of the sample, where force is applied). The magenta labels in  \ref{fig:PressureScaling00} indicate the range of layers over which the summation is performed. 

The only free parameter that needs adjustment for the proper description of $P_{xy}(t)$ is $\Omega$. We used for it the same value in numerical calculations as used elsewhere, i.e. $\Omega$=24.82/ps.

\subsection{Virial stress propagation.}
\label{VirialStress}

\begin{figure}[!ht]
	\centering
	\includegraphics[scale=1.0]{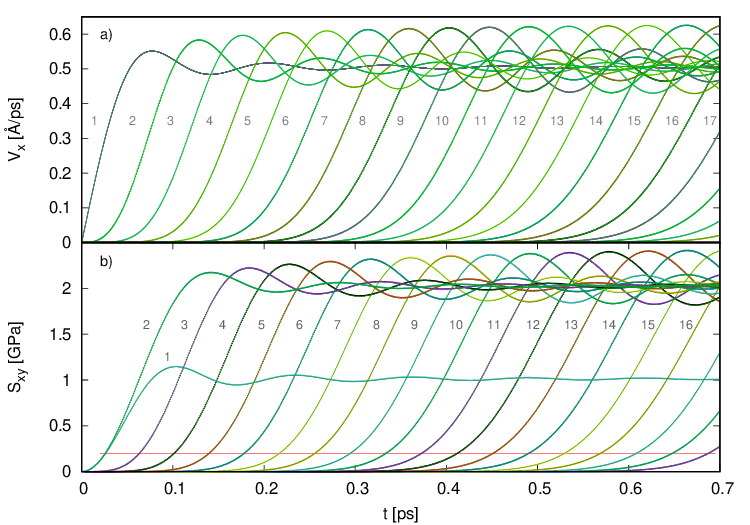}
	\caption{Comparison of analytical results obtained with Bessel functions (green lines in the background) with MD simulations (dots drawn on top of these) for velocity (a) and virial stress (b) of layers. Labels denote layers number, from the uppermost 1. The surface Heaviside impulse pressure applied is 2000 MPa, the sample is with N=33 layers, and $n$=2.00. Here $\Omega$ is 24.82/ps, the same as in  Fig. \ref{fig:PressureScaling00}.
		The virial pressure (force) is computed with Pater's Eq. \ref{eq:fn}, and the velocity of layers with Eq. \ref{eq:vn}.
		The red horizontal line in b) is drawn at 0.1 of the applied surface pressure.
	}
	\label{fig:VBesselHEAVYZ00}
\end{figure}

\begin{figure}[!ht]
	\centering
	\includegraphics[scale=0.25]{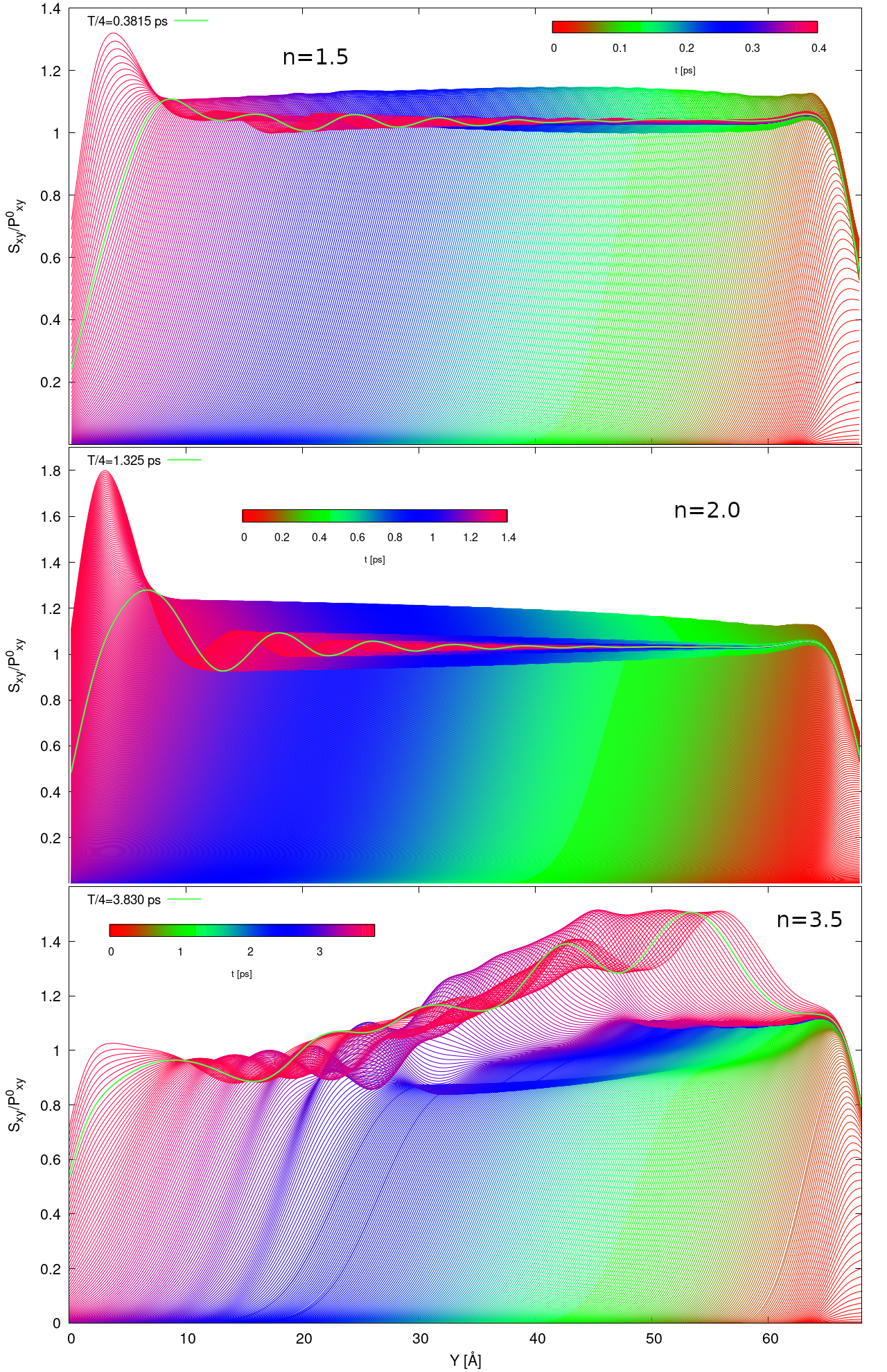}
	\caption{Comparison of the dynamics of stress penetration into the sample interior for three values of non-linearity parameter n: 1.5, 2.0, and 3.5, when the surface applied pressure is 1000 MPa. The upper layer is at around Y=68 {\AA}. Green lines are drawn for $t=T_N/4$. For n=2, \href{http://nanophysics.pl/movie01.mp4}{\textbf{this movie}} illustrates better the process of stress penetration.
	}\label{fig:planes00}
\end{figure}


At very short times, less than the time needed for the sound to pass the distance between two layers, the first only contribution, $f_1$ of Eq.  
\ref{eq:sumfn} is sufficient for the approximation of surface pressure. In that time range, MD results on $\Delta X(t)$ have a parabolic $t$-dependence, in perfect agreement with the Pater analytical results. That is shown by the red dots in the inset of \ref{fig:aHeavy_150}, computed for $n$=2 in the same way (by using 
\ref{eq:un} for the displacement).

The velocity of layers is computed with 
\ref{eq:vn} and compared with the MD simulation results in \ref{fig:VBesselHEAVYZ00} a), for a sample with N=33, for $n$=2, under a surface pressure of 2000 MPa. First, by green lines, the results of computation with Bessel functions are drawn, for layers from the uppermost, labeled as layer 1. On the top of these lines, with points, the results of simulations are shown. There are only two parameters needed for obtaining a coincidence between analytical and MD simulation curves. These are $\Omega$ in 
\ref{eq:vn} that determines the time scale, and the amplitude of velocity, both in agreement with all the scaling relationships discussed so far. 

For the first, uppermost layer velocity can be written as an infinite sum 
\ref{eq:vn}: $v_1 = J_{1}(2\theta) + 2\cdot J_3(2\theta)+2\cdot J_5(2\theta) + ...$. However, for a sample with the number of layers equal to N (N $\ge$ 2) summation must be limited to $J_k$, where k=4(N-1)-1. Hence, the oscillations of the first layer are shown by green lines for samples with N from 2 to 20.

While in \ref{fig:PressureScaling00} we demonstrate that Pater's equations allow us to reconstruct surface pressure as a function of time, in  \ref{fig:VBesselHEAVYZ00} b) we show that virial stress for layers inside of the sample volume can also be computed with these equations, with a very good agreement between analytical results and MD simulation data. The labels indicate the layers, as in figure a). 
There are two important details here. 1) In LAMMPS, the entire force is reported as acted on layers, while Pater equations provide force acting on one side of the layer, only. Therefore, to compare the results of LAMMPS simulations and Pater's, we need to take a sum of two results from Pater for neighboring planes. 2) There is an exception from the rule 1). In the case of the uppermost layer, the force in LAMMPS simulations acts on the layer from one side only (since there is no layer above it). The same is in the case of the lowermost layer. Therefore in these cases, the Pater result for one layer only ought to be used. This explains why the virial stress for the layer labeled as 1 (the uppermost) in \ref{fig:VBesselHEAVYZ00} b) is about 2 times smaller than that for the remaining layers. 

The amplitude of velocity used in \ref{fig:VBesselHEAVYZ00} a) that best fits the MD simulation results is 0.505 {\AA}/ps. That amplitude is given by $F_1/\rm{m}\Omega$ 
\ref{eq:vn}. With the mass of the Fe atom $9.273\cdot 10^{-26}$kg, surface area per atom 5.49 {\AA}$^2$, and $\Omega$=24.82/ps, the amplitude of velocity is found to be 0.48 {\AA}/ps.

Calculations like these, with the use of equations \ref{eq:un}-\ref{eq:fn}, have, however, a limited applicability and can be used for not more than a few tens of layers. Computing Bessel functions requires the summation of a large number of terms, which finally leads to computational instabilities when either the number of layers and/or time becomes too large. In our case, for computing Bessel functions,  we used \texttt{Math::Cephes} Perl libraries.

While \ref{fig:VBesselHEAVYZ00} b) shows the virial stress as a function of time for different layers, in \ref{fig:planes00} we draw virial stress $S_{xy}$ profiles across the sample length in the Y-direction, for increasing time moments, with the time separation of 0.001 ps. The curves of $S_{xy}(y,t)$ have been obtained by applying an \texttt{approximate cubic splines} method (available in Gnuplot) to a discrete data set of points (one point for one layer, with a layer's separation of 2.06 {\AA}).
Pressure is applied at the surface of the sample (with a Y-coordinate of around 68 {\AA}), and the stress front penetrates the interior and reaches the opposite sample side at Y=0, where it is reflected, and the reflected wave interferes with the incoming wave in a constructive way causing an additional increase of virial stress. The light-green lines show the stress profiles at approximately $t=T_N/4$.

The results shown here suggest that the amplitude of oscillations in quantities like $v_x(t)$ or $S_{xy}(t)$, or as a function of the Y-coordinate, grows up with time or with spacial distance. That is a typical result characteristic of solutions to the model of a chain of masses and springs. A natural question arises is the observed increase of amplitude of oscillations going to continue increase with time/distance?
This question can not be resolved by performing numerical modeling by using Bessel functions as provided by Pater equations - the range where accurate approximations of these functions is available is too narrow. We did preliminary MD simulations, performed on very long samples (up to 1000 layers; 0.2 $\mu$m in size), but still were not able to find clear conclusions. The nature of oscillations changes at long times (distances). The phenomenon is related to the harmonic case as well, when $n=2$, but the effects are stronger for $n$ increasing well over 2 and for increasing values of pressure at the surface. Some of our preliminary studies suggest, however, that under high pressures, a catastrophic, chaotic scenario of stress propagation is realized, i.e., the wave front changes abruptly its own shape. On another hand, there are theoretical arguments that the amplitude of similar oscillations decreases with time/distance \cite{Lowell}. This subject requires separate systematic research, theoretical and based on MD simulations. Perhaps the new experimental techniques available \cite{PNAS} could also provide physical evidence for the nature of these phenomena.

\clearpage

\subsection{Sound velocity.} 
\label{SoundVelocity}

\begin{figure}[!ht]
	\centering
	\includegraphics[scale=1.0]{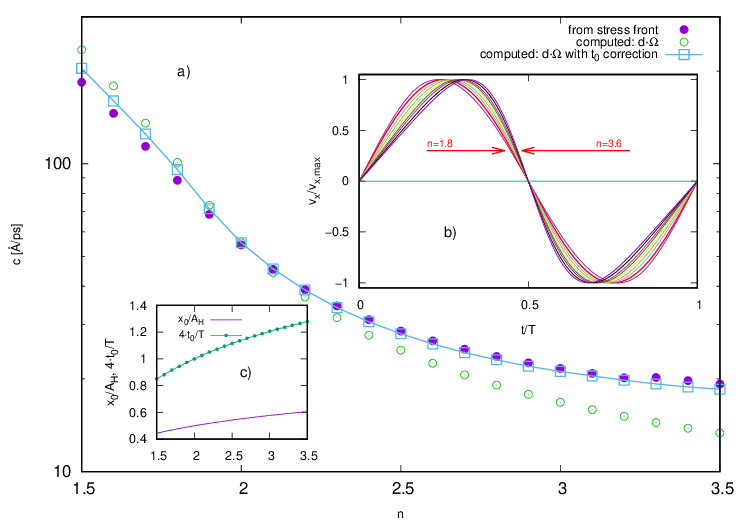}
	\caption{
		a) Speed of sound is determined as a ratio of the distance between 
		stress curves for layers 2-nd and 33-rd, like those in \ref{fig:VBesselHEAVYZ00} b), and the time needed for the travel of pressure wave between these layers. It is shown with full circles.  
		b) Change with $n$ of the position of velocity maximum in a 2-layer case has been used to find out the ratio of $t_0/T$, as shown in c). The speed of sound corrected by the factor $1/(2-4t_0/T)$ is shown in a) with empty squares.
	}
	\label{fig:sound01}
\end{figure}

A peculiar property of Pater's solutions is an infinite speed of sound in the limit of very small displacements, as noticed by Schr\"{o}dinger \cite{Erwin},\cite{Muhlich}. The displacement of the m-th layer, $u_m$, is proportional to $J_{2m}(2\theta)$, which for small arguments can be approximated by $u_{m} \sim 1/(2m)! \cdot (\Omega t)^{2m}$. Hence, $u_{m}$ is non-zero (albeit it may remain very small) for any $m$ and for any arbitrarily small time $t>0$.

It is difficult to formulate a precise, rigorous criterion for determining the time when stress entering the volume starts to increase at a position of a given layer, or when displacement starts to change. That increase in time is initially slow. We are able to show that $u_m(t)$, as determined from MD simulations, can be accurately described by $(\Omega t)^{2m}$ functions of time, however for not more than 4 layers, only. 

We can determine the speed of sound by using analytical equations of Pater and numerical methods of finding out when a certain threshold displacement of $u_m$ is reached for various layers, to determine from that the speed of sound $c$. That method gives results on $c$ dependent on the value of the critical threshold assumed. However, that dependence is not strong and in the limit of large thresholds (when $u_m$ is of the order of 1) the well anticipated limit of a classical infinite chain of masses is approached.

The red horizontal line in \ref{fig:VBesselHEAVYZ00} b) is drawn at 0.1 of the applied surface pressure and the points of crossing it by $S_{xy}(t)$ curves have been used to determine the speed of sound (this is a transverse sound component, with propagation in direction [111]). Full circles in \ref{fig:sound01} a) show sound velocity determined in that way, i.e. from the criterion of reaching 0.1 of $P^0_{xy}$, when pressure applied is 1000 MPa. 

In the case of linear theory (when $n$=2), it follows from dispersion relations for propagation of waves in a periodic medium that sound velocity $c$ is related to angular frequency $\Omega$, and to the distance $d$ between layers, $c=d\cdot \Omega$. In our case, $d=a/\sqrt{3}$, where lattice constant $a$ is 3.56 {\AA}. For comparison, empty circles in \ref{fig:sound01} a) show the speed of sound computed with that equation. It does reproduce perfectly the speed determined for $n$=2, but for other values of the non-linearity parameter deviations are significant.

Another method of determining the speed of sound is based on measurements of the slope of the surface pressure normalized by the amplitude of applied pressure, $P_{xy}(t)/P^0_{xy}$, during times between around 0.1$T_N$ and $T_N$/4. It is based on the assumption that the front of pressure wave does not change with time, and therefore a uniform stress within the sample volume penetrated already by the wave is achieved. It leads us to the relation:

\begin{equation}
	c=  \frac{d(P_{xy}(t)/P^0_{xy})}{dt} \cdot (N-1)\cdot d.
	\label{eq:soundP}
\end{equation}

That method gives us results on $c(n)$ nearly identical to those determined with the first method when $n$<2, with some small and interesting differences for larger $n$. It is simpler and faster than the first one. However, it must be used with caution in situations of non-linear materials, unless we understand well the mechanisms of wave propagation in a studied medium.

We think that the velocity of stress propagation is influenced largely by the asymmetry of the potential well, as shown in \ref{fig:P07}. The time needed for a particle to move on the left of the potential minimum is different from the time for traversing the right part: for $n$>2 the time on the right is shorter than the time on the left. This asymmetry is well seen in \ref{fig:sound01} b), where velocity normalized by the period of oscillations $T$ for a 2-layer case is shown for a range of $n$-values. Hence, we argue that the time on the right determines the speed of sound. 

The point of potential minimum is marked as $(x_0,t_0)$ in \ref{fig:P00}. Position $x_0$ is known from 
\ref{eq:AMP}. Since we have no analytical expression on $x(t)$, we cannot  provide an exact formula on $t_0$. One of the possible methods of determining $t_0/T$ is to find it from the position of maximum velocity, as in \ref{fig:sound01} b). The results on dependence of $4\cdot t_0/T$ on $n$ are shown in  \ref{fig:sound01} c), together with $x_0/A_H$.  

When we correct the sound speed computed from equation $c=d\cdot\Omega$ by a factor $1/(2-4t_0/T)$ (this is like assuming that the time $T/2-t_0$ should replace $T/4$ on the period of harmonic oscillator), the results become very close to those determined from the measurements of the front wave, as shown by empty squares and the blue curve (for guiding eyes only), in \ref{fig:sound01} a).

The determined two-layer oscillation frequency $\omega$ in the displacement method for n=2 is 26.9 $\cdot 10^{12}$/s, and the speed of the transverse sound is 55.404 {\AA}/ps. Both these values are about 2 times higher than those found by us for steel 310S, when Bonny EAM inter-atomic potential is used, where we have 14.6 $\cdot 10^{12}$/s and 30.29 {\AA}/ps, respectively. The results of MD simulations for steel are in agreement with the ab-initio computations \cite{Muller} and with experimentally determined phonon energies in Fe \cite{Zarestky} or steel \cite{Danilkin}, \cite{Urbassek}. That means that the parameter $\epsilon_0$ used as the amplitude of the potential energy well ought to be rescaled down about 4 times to reproduce properly the results found in steel.

\section{Summary and conclusions.}

The method has been described for creating a table-style non-linear inter-atomic potentials between NN atoms in FCC crystals for the use in MD simulations in LAMMPS. Exact analytical formulas were found on the period of oscillations for a system of two-layer crystallographic planes, when one of them is initially displaced, and approximate equations are provided for the case when Heaviside-type pressure is applied. Scaling relations are given between quantities such as average displacement, velocities, and forces acting on the planes. The results of that analysis helped us understand the dynamics of stress propagation into the bulk of crystal, consisting of many crystallographic planes. 

We used analytical Pater's model of the chain of masses and springs to find out that it describes perfectly well the data obtained by MD simulations in the case of harmonic inter-atomic potential. The model implies the existence of strong oscillations of stress, displacement and velocities as a function of time and position within the volume of elastic medium. We found also that in non-linear cases Pater's model remains qualitatively valid as well, since a quasi-linear relation between the wave vector and frequency of oscillations is preserved. That allowed us to determine that the speed of sound is a strongly decreasing function of $n$. In case, however, when non-linearity parameter $n$ becomes larger than 2 and/or in case of large applied pressures, oscillations may become unstable and chaotic. The subject of stability of propagating waves of stress requires separate systematic research. 

Oscillations like those of $v_x(t)$, $S_{xy}(t)$, etc., can easily be misinterpreted as a result of typical time dependencies found when pressure is applied abruptly and energy losses are present. Oscillations in such cases can be observed in surface pressure $P_{xy}(t)$ as well (some of the results on $P_{xy}(t)$ presented in \cite{Koziol2} have been misinterpreted by us).

Similar oscillations of various quantities are often observed in MD simulations, either due to some justified physical origin or caused by an unintentional mistake in the methodology of performing the simulations. Therefore, the significance of Pater's model is in explaining that in a case of perfect harmonic inter-atomic potential and no energy losses, we still ought to observe oscillations in certain conditions that are related to the discrete nature of atoms and their periodic arrangement in crystals. 

The developed methodology of performing MD simulations and subsequent data analysis allowed us to achieve unprecedented accuracy, with results in very good agreement with the theoretical, analytical model for n=2.

An unresolved question is how adding more degrees of freedom (in particular, allowing the movement of atoms in the Z-direction) will change the dynamics of stress propagation? What will be the effect of chaotic movement due to the finite temperature on the propagation of waves and the stability of the crystal lattice? Will finite temperature change the effective inter-atomic potential, and how will it influence results of the model of chain of masses and springs? 


\bibliographystyle{iopart-num}
\bibliography{anharmonicV2}

\providecommand{\newblock}{}
\begin{thebibliography}{10}
\expandafter\ifx\csname url\endcsname\relax
  \def\url#1{{\tt #1}}\fi
\expandafter\ifx\csname urlprefix\endcsname\relax\def\urlprefix{URL }\fi
\providecommand{\eprint}[2][]{\url{#2}}

\bibitem{PNAS}
Holstad T~S, Dresselhaus-Marais L~E, Ræder T~M, Kozioziemski B, van Driel T,
  Seaberg M, Folsom E, Eggert J~H, Knudsen E~B, Nielsen M~M, Simons H, Haldrup
  K and Poulsen H~F 2023 {\em PNAS\/} {\bf 120(39)} 1--4

\bibitem{Koziol2}
Kozio{\l} Z 2022 {\em Modelling Simul. Mater. Sci. Eng.\/} {\bf 30} 065010

\bibitem{Ashcroft}
Ashcroft N~W and Mermin N~D 1976 {\em Solid State Physics\/} 1st ed (New York:
  Saunders College Publishing)

\bibitem{Kittel}
Kittel C 2005 {\em Introduction to solid state physics\/} 8th ed (New York:
  Wiley)

\bibitem{Katsnelson}
Katsnelson M~I 2005 {\em Lattice dynamics: anharmonic effects, in: Encyclopedia
  of Condensed Matter Physics\/} 1st ed (Amsterdam: Elsevier)

\bibitem{Cowley}
Cowley R~A 1968 {\em Rep. Prog. Phys.\/} {\bf 31} 123

\bibitem{Complexity}
D C, R S and R F 2003 {\em Complexity\/} {\bf 8(3)} 19--30

\bibitem{Herrera}
Herrera M, Antonsen T~M, Ott E and Fishman S 2013 {\em Phys. Rev. A\/} {\bf 87}
  041603(R)

\bibitem{Pathak}
Pathak K~N 1965 {\em Phys. Rev.\/} {\bf 139(5A)} A1569

\bibitem{Lowell}
Lowell S~C 1970 {\em Proc. R. Soc. Lond. A\/} {\bf 318} 93--106

\bibitem{Hutereau}
Hutereau M, Banks P~A, Slater B, Zeitler J~A, Bond A~D and Ruggiero M~T 2020
  {\em Phys. Rev. Lett.\/} {\bf 125} 103001

\bibitem{Nucera}
Nucera C and di~Scalea F~L 2014 {\em Journal of Sound and Vibration\/} {\bf
  333} 541--554

\bibitem{Hu}
Hu S~Q, Chen D~Q, Zhang S~J, Liu X~B and Meng S 2022 {\em Materials Today
  Physics\/} {\bf 27} 100790 ISSN 2542-5293

\bibitem{Hoegen}
von Hoegen A, Mankowsky R, Fechner M, Forst M and Cavalleri A 2018 {\em
  Nature\/} {\bf 555} 79

\bibitem{Bonny}
Bonny G, Castin N and Terentyev D 2013 {\em Modelling and Simulation in
  Materials Science and Engineering\/} {\bf 21} 085004

\bibitem{Artur}
B\'{e}land L, Tamm A, Mu S, Samolyuk G, Osetsky Y, Aabloo A, Klintenberg M,
  Caro A and Stoller R 2017 {\em Comput. Phys. Comm.\/} {\bf 219} 11--19

\bibitem{Grado}
Grado-Caffaro M and Grado-Caffaro M 2011 {\em Journal of Physics and Chemistry
  of Solids\/} {\bf 72} 957--960

\bibitem{Carre}
Carré A, Ispas S, Horbach J and Kob W 2016 {\em Computational Materials
  Science\/} {\bf 124} 323--334

\bibitem{LAMMPS}
Plimpton S 1995 {\em J. Comp. Phys.\/} {\bf 117} 1

\bibitem{Hellman}
Hellman O, Abrikosov I~A and Simak S~I 2011 {\em Phys. Rev. B\/} {\bf 84}
  180301(R)

\bibitem{Erwin}
Schr{ö}dinger E 1914 {\em Ann. Phys.\/} {\bf 349(14)} 916--934

\bibitem{Muhlich}
U M, BE A and dell’Isola F 2021 {\em Mathematics and Mechanics of Solids\/}
  {\bf 26(1)} 133--147

\bibitem{Pater}
de~Pater A~D 1974 {\em Vehicle System Dynamics: International Journal of
  Vehicle Mechanics and Mobility\/} {\bf 3:3} 123--140

\bibitem{Lui}
H L~C and F H~T 2013 {\em Phys. Rev. B\/} {\bf 87} 121404

\bibitem{Tan}
Tan P, Han W, WJ, Zhao, Wu Z, Chang K, Wang H, Wang Y, Bonini N, Marzari N,
  Savini G, Lombardo A and Ferrari A 2012 {\em Nat. Mat.\/} {\bf 11} 294

\bibitem{Koziol}
Kozio{\l} Z, Gawlik G and Jagielski J 2019 {\em Chin. Phys. B\/} {\bf 28}
  096101

\bibitem{Belland}
E S~R, A T, K B~L, D S~G, M S~G, A C, V S~L, N O~Y, A A, M K and Y W 2016 {\em
  J. Chem. Theory Comput.\/} {\bf 12(6)} 2871 -- 2879

\bibitem{Butcher}
Butcher J~C 2008 {\em Numerical Methods for Ordinary Differential Equations\/}
  (New {Y}ork: John Wiley {\&} Sons)

\bibitem{Euler}
Euler N 1997 {\em Journal of Nonlinear Mathematical Physics\/} {\bf 4(3-4)}
  310--337

\bibitem{Amore}
Amore P and Fern\'{a}ndez F~M 2005 {\em Eur. J. Phys.\/} {\bf 26(4)} 589

\bibitem{Harko}
Harko T, Lobo F~S~N and Mak M~K 2013 {\em Journal of Pure and Applied
  Mathematics: Advances and Applications\/} {\bf 10(1)} 115--129

\bibitem{Znojil}
Znojil M and Semor\'{a}dov\'{a} I 2018 {\em Modern Physics Letters A\/} {\bf
  33(38)} 1850223

\bibitem{Barreto}
Barreto R, Carusela M~F and Monastra A~G 2017 {\em J. Stat. Mech.\/}  103201

\bibitem{Pankov}
Pankov A 2019 {\em Discrete and Continuous Dynamical Systems - S\/} {\bf 12(7)}
  2097--2113

\bibitem{Kashchenko}
Kashchenko S 2020 {\em Communications in Nonlinear Science and Numerical
  Simulation\/} {\bf 91} 105436

\bibitem{FPU}
Fermi E, Pasta P, Ulam S and Tsingou M 1955

\bibitem{Landau}
Landau L~D and Lifshitz E~M 1976 {\em Mechanics: Volume 1 (Course of
  Theoretical Physics)\/} 3rd ed (Oxford: Butterworth-Heinennan)

\bibitem{Abramowitz}
Abramowitz M and Stegun I (eds) 1970 {\em Handbook of Mathematical Functions
  with Formulas, Graphs, and Mathematical Tables\/} 9th ed (New {Y}ork:
  {D}over)

\bibitem{Lawden}
Lawden D~F 1989 {\em Elliptic Functions and Applications. Applied Mathematical
  Sciences\/} vol~80 (New {Y}ork: Springer-Verlag) ISBN 1441930906

\bibitem{Whittaker}
Whittaker E~T 1964 {\em A Treatise on the Analytical Dynamics of Particles and
  Rigid Bodies\/} 4th ed (Cambridge: Cambridge University Press)

\bibitem{Folland}
Folland G~B 1992 {\em Fourier analysis and its applications, The Wadsworth and
  Brooks/Cole Mathematics Series\/} 1st ed (Pacific Grove: Wadsworth and
  Brooks/Cole Advanced Books and Software)

\bibitem{Herman}
Herman R~L 2016 {\em An Introduction to Fourier Analysis\/} 1st ed (CRC Press)

\bibitem{Yang}
Yang J and Komvopoulos K 2020 {\em Int. J. Solids Struct.\/} {\bf 193--194}
  98--105

\bibitem{Zhou}
Zhou M 2003 {\em Proc. R. Soc. Lond. A\/} {\bf 459} 2347--2392

\bibitem{Subramaniyan}
Subramaniyan A~K and Sun C 2008 {\em Int. J. Solids Struct.\/} {\bf 45}
  4340--4346

\bibitem{Zimmerman}
Zimmerman J, III E~W, Hoyt J, Jones R, Klein P and Bammann D 2004 {\em
  Modelling Simul. Mater. Sci. Eng.\/} {\bf 12} S319

\bibitem{Elder}
Elder R~M, Mattson W~D and Sirk T~W 2019 {\em Chem. Phys. Lett.\/} {\bf 731}
  136580

\bibitem{Muller}
Müller M, Erhart P and Albe K 2007 {\em J. Phys.: Condens. Matter\/} {\bf 19}
  326220

\bibitem{Zarestky}
Zarestky J and Stassis C 1987 {\em Phys. Rev. B\/} {\bf 35} 4500

\bibitem{Danilkin}
Danilkin S~A, Fuess H and Wieder T 2001 {\em J. Mater. Sci.\/} {\bf 36} 811

\bibitem{Urbassek}
Urbassek H and Sandoval L 2012 Diffusionless transformations high strength
  steels modelling and advanced analytical techniques {\em Phase
  Transformations in Steels\/} ed Pereloma E and Edmonds D~V (Woodhead
  Publishing)

\end{thebibliography}

\end{document}